\documentclass[]{tCPH2e}

\begin{document}
\doi{10.1080/0010751YYxxxxxxxx}
 \issn{1366-5812}
\issnp{0010-7514}


\markboth{Taylor \& Francis and I.T. Consultant}{Contemporary Physics}

\title{\itshape Optical antennas and plasmonics}

\author{Q-Han Park\thanks{ Email: qpark@korea.ac.kr}
\\\vspace{6pt}
{\em{Department of Physics, Korea University, Seoul, 136-701,
Korea}; } }  

\maketitle

\begin{abstract}
Optical antenna is a nanoscale miniaturization of radio or microwave antennas that is also governed by the rule of plasmonics.
We introduce various types of optical antenna and make an overview of recent developments in optical antenna research.
The role of local and surface plasmons in optical antenna is explained through antenna resonance and resonance conditions for
specific metal structures are explicitly obtained.
Strong electric field is shown to exist within a highly localized region of optical antennas such as antenna feed gap
or apertures. We describe physical properties of field enhancement in apertures(circular and rectangular holes) and
gaps(infinite slit and feed gap), as well as experimental techniques measuring enhanced electric vector field.
We discuss about analogies and differences between conventional and optical antennas with a projection of future developments.

\bigskip
\bigskip

\end{abstract}

\section{Introduction}
Antennas are ubiquitous in modern day communications and come in various forms. In early days, from the Marconi's first field
test of antennas until World War II, most antennas were of wire type which was used to transmit or receive radio waves.
Developments of microwave technology during and after the War introduced new types of antenna (apertures, microstrips,
phased-array radars, etc.)\cite{Balanis} and shrunk the size of antenna as size depends on the wavelength of an electromagnetic wave.
In the simplest form, antenna is a quarter wavelength long metal pole mounted on a conducting plate
which is known as the monopole antenna.
The typical size of antennas for radio or microwave communications ranges from hundred meters to a few milimeters.
The shift of interest to the higher frequency band of electromagnetic waves in communication, undoubtably optical fiber communication is
one good example, and the advance in nanotechnology and optical science raise one natural question -
is it possible to make ``optical antennas"?
At optical frequencies, according to the quarter wavelength resonance condition, the size of antenna may be only about hundred nanometers !
But can we simply shrink down the antenna in proportion to the wavelength to make optical antennas work? In optical fiber
communication, waves are already guided by a fiber and there seems to be no need for antennas which are used for wireless communications.
On the other hand, in wireless optical communication, optical components such as mirror and lens are used to receive optical signals
instead of subwavelength optical antennas as we can see in the case of a telescope.
Then, what can we really do with optical antennas ?
\\

During the past decade, there has been a surge of research activity on the optical properties of metal nanoparticles and metals
with structures at the nanometer scale. The advances in fabrication technique enabled construction of metallic nano structures in various forms.
In particular, an optical analogue of monopole antenna - a metal pole of length about 100 nm mounted on a nearfield scanning
optical microscopy(NSOM) tip as illustrated in Fig. 2b has been constructed \cite{Frey1,Frey2,Taminiau1,Taminiau2,Taminiau3}.
Antenna resonance and field localization near an optical monopole antenna have been demonstrated by using single fluorescent
molecules \cite{Taminiau1}. These efforts begun with a realization that an optical antenna can be used to increase the efficiency
of a near-field optical probe that has a spatial resolution well below the diffraction limit \cite{Grober}.
The subwavelength resolution below the diffraction limit was made possible through the invention of the metal coated, tapered fiber
probe with an aperture of size much smaller than wavelength \cite{Lewis,Pohl,Betzig}. Subwavelength size apertures, however, suffer
from a poor coupling efficiency of nearby light since it is known that the coupling efficiency scales as $(a/\lambda)^4 $,
where $a$ is the diameter of a hole and $\lambda $ is wavelength.
Optical antennas can increase the coupling efficiency by focusing light onto an aperture region.
\\

Another important antenna  type is a dipole antenna which is made of two monopole rods separated by a small
gap. The conducting plate of a monopole antenna provides the image of a monopole rod and subsequently
the monopole antenna is in fact equivalent to the dipole antenna. Optical dipole antennas have been used widely
for nearfield optical probes. On resonance, strong field enhancement in the dipole antenna gap
was observed with nanometer-scale gold dipole antennas \cite{Muhlschlegel,Ghenuche}. Similar enhancement was also obtained
with a nanometer-scale bowtie antenna \cite{Schuck}. Bowtie antenna, a variant of dipole antenna made of two triangle shaped metal pieces,
has been used widely for an optical antenna \cite{Grober,Schuck,Farahani1,Farahani2,Guo,Fischer,Huang,Fromm,Wang,Sundaramurthy}.
Strong electric field with an enhancement factor several orders of magnitude can modify the emission of light by a molecule.
Increased fluorescence of molecules near optical antennas has been reported \cite{Taminiau1,Kuhn,Bharadwaj1,Bharadwaj2,Tam,Bakker}.
Moreover, spectroscopic techniques such as surface-enhanced Raman scattering(SERS) \cite{Gersten}, in the presence of a highly localized
strong electric field, allowed optical detection and spectroscopy of a single molecule \cite{Nie,Kneipp,Xu,Felidj,Rogobete,Taminiau}.
Single molecule detection has been also demonstrated using local fields resonantly enhanced by monopole antennas mounted
on a NSOM tip \cite{Taminiau1,Taminiau2,Taminiau3} and applied to biological nano-imaging \cite{Frey2,Garcia-Parajo}.
In this regard, optical antenna is more of a probe for the communication with the nanoscopic world than being a tool for
human-to-human communications.
\\

\begin{figure}
\begin{center}
\begin{minipage}{100mm}
\subfigure[]{
\resizebox*{5cm}{!}{\includegraphics{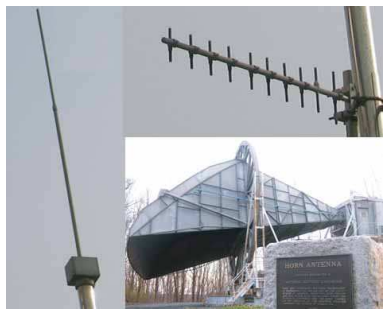}}}%
\caption{Various antennas: monopole antenna (left), Yagi-Uda antenna consisting of dipole antenna arrays(right top)
and the horn antenna(right bottom) at Bell Labs, Holmdel, NJ that Penzias and Wilson in 1965 used to discover the
cosmic microwave background radiation, the evidence of Big Bang theory.  }%

\end{minipage}
\end{center}
\end{figure}

The enhancement of electric field in an optical dipole antenna is caused essentially by two factors: surface charge and resonance.
Current on an antenna surface which is induced by an incoming light charges the feed gap of a dipole antenna and subsequently establishes
a strong electric field in the gap. Gap charging becomes resonantly enhanced when the antenna geometry meets the resonance condition;
that is, the length of a dipole antenna is an integer multiple of half wavelength. The resonance condition, which holds
for radio and microwave antennas, does not apply directly to optical antennas.
Metal nanorods or wires excited by light show resonance characteristics which are different from the perfectly conducting
case \cite{Imura,Novotny,Hanson,Ditlbacher,Neubrech,Laroche,Payne,Crozier,Schider}.
At optical frequencies, metal is a highly dispersive material with finite conductivity. Electron gas in a
metal can couple to light in the form of a propagating surface wave also called surface plasmon polariton(SPP)\cite{Ritchie,Raether}
or local plasmons in the case of nanoparticles \cite{Kelly}.
Plasmons, either SPP or local plasmons, exhibit salient resonance features. Due to the excitation of local plasmons, the optical properties
of metal nanoparticles for instance are determined by the shape, size and dielectric environment of a metal particle \cite{Kelly}.
Recent development of nanofabrication techniques enabled construction of a variety of metal structures at the subwavelength scale
and opened the research area called ``plasmonics", a subfield of nanophotonics studying the manipulation of light coupled to electrons
at the nanoscale \cite{Barnes,SAMaier,Lal}. The dispersion relation of SPP and the local plasmon resonance complicate
the resonance feature of optical antennas and modify significantly other features of a conventional antenna.
Roughly speaking, one can not simply shrink down a radio antenna to make an optical antenna.
The properties of optical antennas are still under the intensive study. We may say that research efforts to relate plasmonics
with subwavelength optical antenna have just begun.
\\

The slot antenna is a complementary structure of a dipole antenna formed by cutting a narrow slot in a large piece of metal plate.
The Babinet's principle suggests that the slot antenna works equally well as its complementary dipole and
various shapes of an aperture can be applied to make optical antennas.
The transmission of electromagnetic wave through a subwavelength size hole is an old subject dating back to
World War II \cite{Bethe,Bouwkamp}, but an optics version of the problem with a new element, plasmon, has ignited
an explosive interest \cite{Ebbesen,Genet} which eventually led to the birth of plasmonics.
Rigorous studies have been made on the transmission of light through a single circular hole
\cite{Degiron1,Degiron2,Zakharian,Alaverdyan,Rindzevicius,Prikulis,Sepulveda,Popov,Abajo} or
other types of apertures such as C-shape \cite{Shi,Matteo,Sun,Lee}.
Bowtie shape aperture antennas have received much attention in terms of their capability to gain the strong field confinement and
enhancement \cite{Jin,Ishihara,Wang1,Wang2,Guo}. Retangular shape aperture is of particular interest
since it is the geometrical shape of a slot antenna. Half wavelength resonance condition applies to the slot antenna of high conductivity
and a half wave slot leads to a resonantly enhanced and aperture confined electric field \cite{Garcia-Vidal,DSKim}.
Aperture type optical antennas have the advantage of light confinement and better control of the radiation pattern
compared to the wire type. This initiated efforts on using bowtie aperture antennas to achieve better subwavelength resolution in near
field imaging or nanoscale lithography \cite{Ishihara,Wang1,Wang2}.
\\

The efficiency of a single optical antenna is generally low and the functionality is limited. Modern microwave antenna technology
created various alternatives to overcome these difficulties including phased-array antenna, reflector antenna, frequency-indendent antenna
and microstrip antenna. These alternatives can be also applied to improve optical antennas. The celebrated enhanced transmission
through a periodic array of holes \cite{Ebbesen} may be attributed to the action of a phased-array optical antenna \cite{Zhang}.
Surrounding a single hole by periodic corrugations, so called the Bull's eye structure, it was demonstrated that the optical aperture antenna
has ability to increase and selectively transmit light with a controlled directivity \cite{Lezec,Degiron,Martin-Moreno}.
The idea of Yagi-Uda antenna \cite{Yagi}, which is a directional antenna using reflectors, has been applied in the optical domain
to obtain a directed emission of single emitters and shape light beams \cite{Taminiau4,Li,Hofmann}. Optical nanostrip antennas have been
studied with an analysis on the slow-plasmon resonance \cite{Bozhevolnyi1,Bozhevolnyi2}.
\\

Optical antennas also open a new realm of application that was not conceived in microwave antennas.
By integrating a resonant optical antenna made of a pair of gold nanorods on the facet of a diode
laser (or a midinfrared quantum cascade laser), a plasmonic laser antenna (or respectively a plasmonic quantum cascade
laser antenna) has been constructed with the ability to confine infrared radiation well below 100 nm \cite{Cubukcu,Yu}.
An array of bowtie antennas was proposed as a new tool to produce coherent extreme ultraviolet light \cite{Kim}. High harmonic generation
was obtained by focusing a femtosecond laser onto a gas inside the bowtie antenna gap.
These efforts demonstrate the exciting potential of optical antennas as a mean of realizing active or nonlinear optical processes
at the nanoscale.
\\

Recent advances in optical antenna theory and application are mostly based on one particular property of an antenna, that is
the strong field enhancement at the nano scale region. Plasmonic resonances add a new twist to the problem of antenna resonance and
field enhancement. As optical antenna is in a rapidly developing stage, in the present paper we will avoid reviewing
optical antennas in detail but will focus on the basic issues of plasmonic resonance and field enhancement which are critical elements of
optical antennas. The paper is organized as follows; section 2 introduces antenna resonance and plasmonic resonances existing in
optical antennas. Plasmon resonance conditions are explained for various metal geometries: flat surface, cylindrical surface
and nanoparticles. Section 3 describes the strong field enhancement in optical aperture type antennas. We explain both resonant
and non-resonant field enhancements existing in holes and slit, and introduce vector field mapping techniques. Finally, in Section 4,
we discuss about analogies and differences between optical antennas and conventional antennas with a projection of future developments.

\begin{figure}
\begin{center}
\begin{minipage}{100mm}
\subfigure[]{
\resizebox*{5cm}{!}{\includegraphics{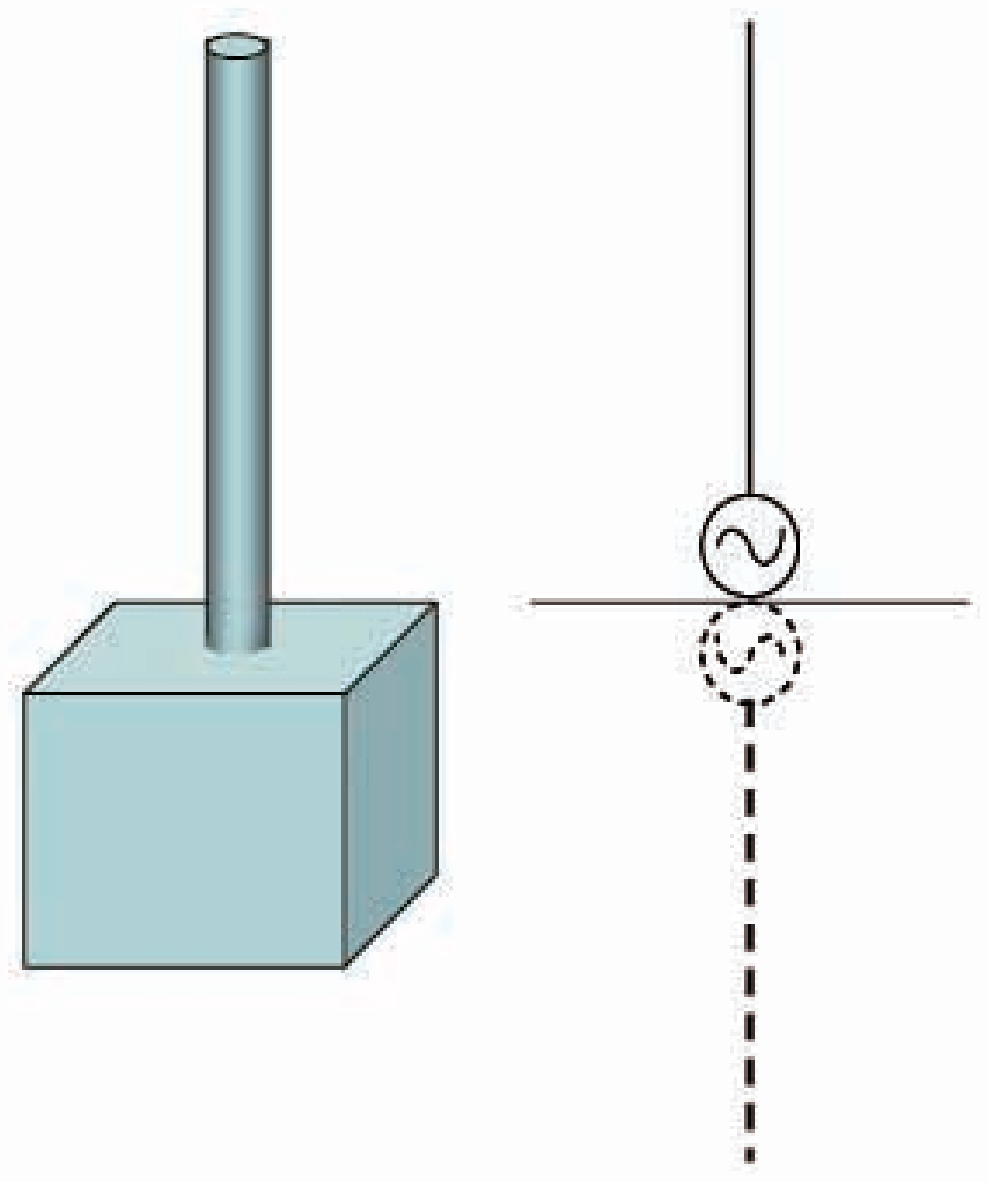}}}%
\subfigure[]{
\resizebox*{5cm}{!}{\includegraphics{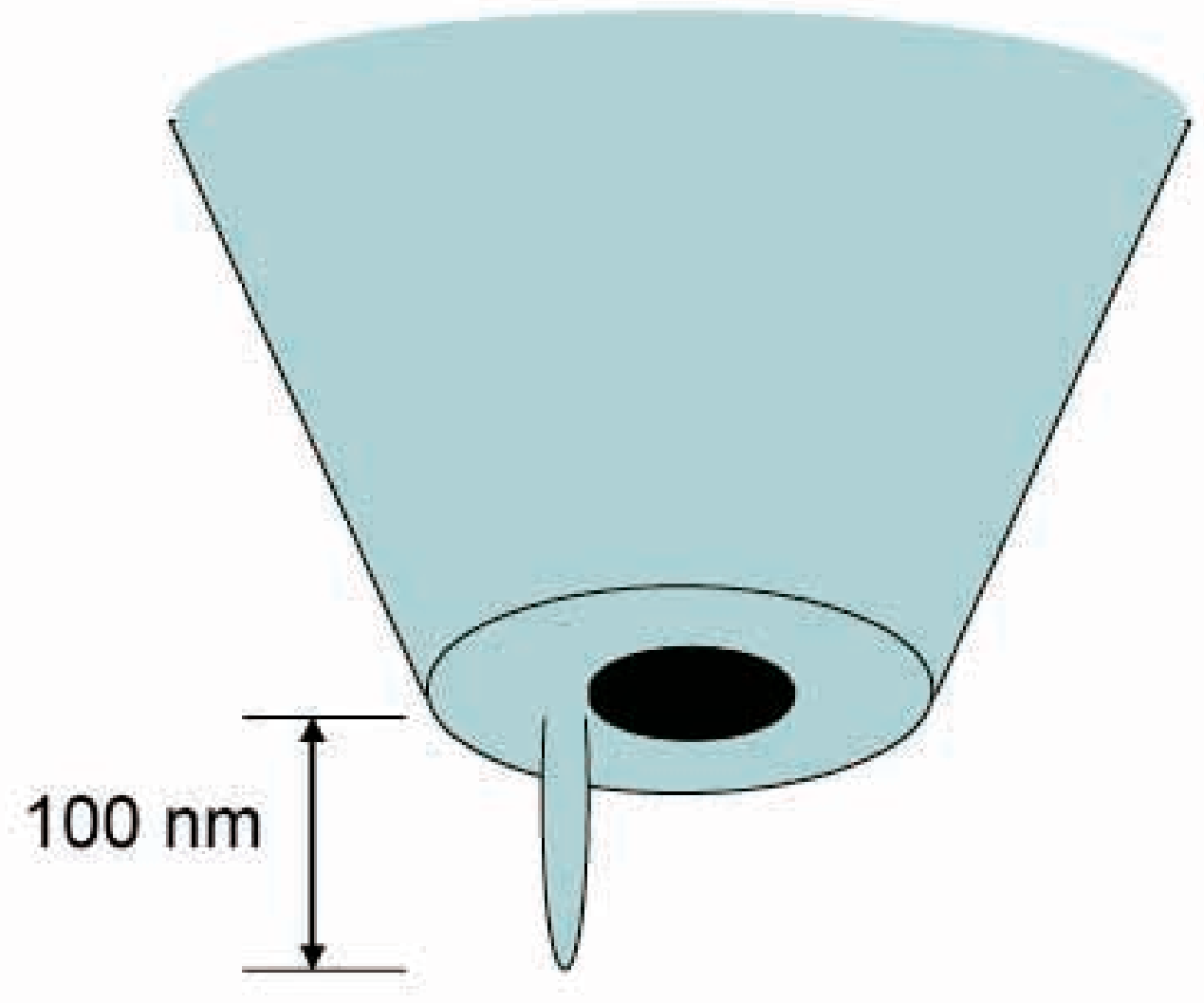}}}%
\caption{Monopole antennas: (a) typical monopole antenna (b) optical monopole antenna on a NSOM tip.  %
Conducting metal mount acts a mirror and pairs up the monopole rod with its image rod to form a dipole antenna.  }%

\end{minipage}
\end{center}
\end{figure}

\section{Antenna resonance and plasmonics}
\subsection{Monopole antenna}

The word antenna as we know was first used by Guglielmo Marconi during his experiment on a wireless transmission
using a metal pole (antenna in Italian means pole). A metal pole mounted on a ground or
a conducting box is known as a monopole antenna. Marconi's monopole antennas receive or transmit radio waves most efficiently when
the pole is one quarter wavelength in length. To understand this, we first recall the radiation mechanism of a wire antenna.
When free electrons on a conducting wire surface are set in motion by the external source, current flows along the surface of a wire.
Electrons are accelerated or decelerated if the the driving source is time harmonic, or if the wire geometry blocks constant flow
of electrons and these accelerated (decelerated) charges, or equivalently time varying currents, create radiation.
To see this more clearly, consider two parallel wires as in Fig. 3a driven by an alternating current source.
When two wires are close, electric field lines, emanating from the positive charges on one wire and ending at the negative charges on
the other wire, are mostly confined within the region between two wires creating little radiation.  If two wires are bent in opposite
directions with open ends as in Fig. 3a, electric field lines becomes circularly distorted. As the traveling current reach to the end,
field lines become pinched and form a closed loop and finally leave the wire. This wire configuration is called dipole antenna and
the radiation pattern is indeed that of a dipole in Fig. 3b. The traveling wave current gets reflected at the end of wire and the reflected
wave, when combined with the incident wave, forms a standing wave current pattern with period $\lambda /2$. Thus, if the total length
of the bending section of wires are integer multiple of $\lambda /2$, the dipole antenna gets resonantly excited increasing
the antenna efficiency. A monopole antenna can be thought of a dipole antenna where one half of a dipole antenna is replaced with
a ground plane at right-angles to the remaining half. The ground plane, when sufficiently large, provides a reflection forming a mirror
image of an antenna  as indicated in Fig. 2a. This explains that the resonance length of a monopole antenna is the integer multiple of $\lambda /4$.

\begin{figure}
\begin{center}
\begin{minipage}{100mm}
\subfigure[]{
\resizebox*{5cm}{!}{\includegraphics{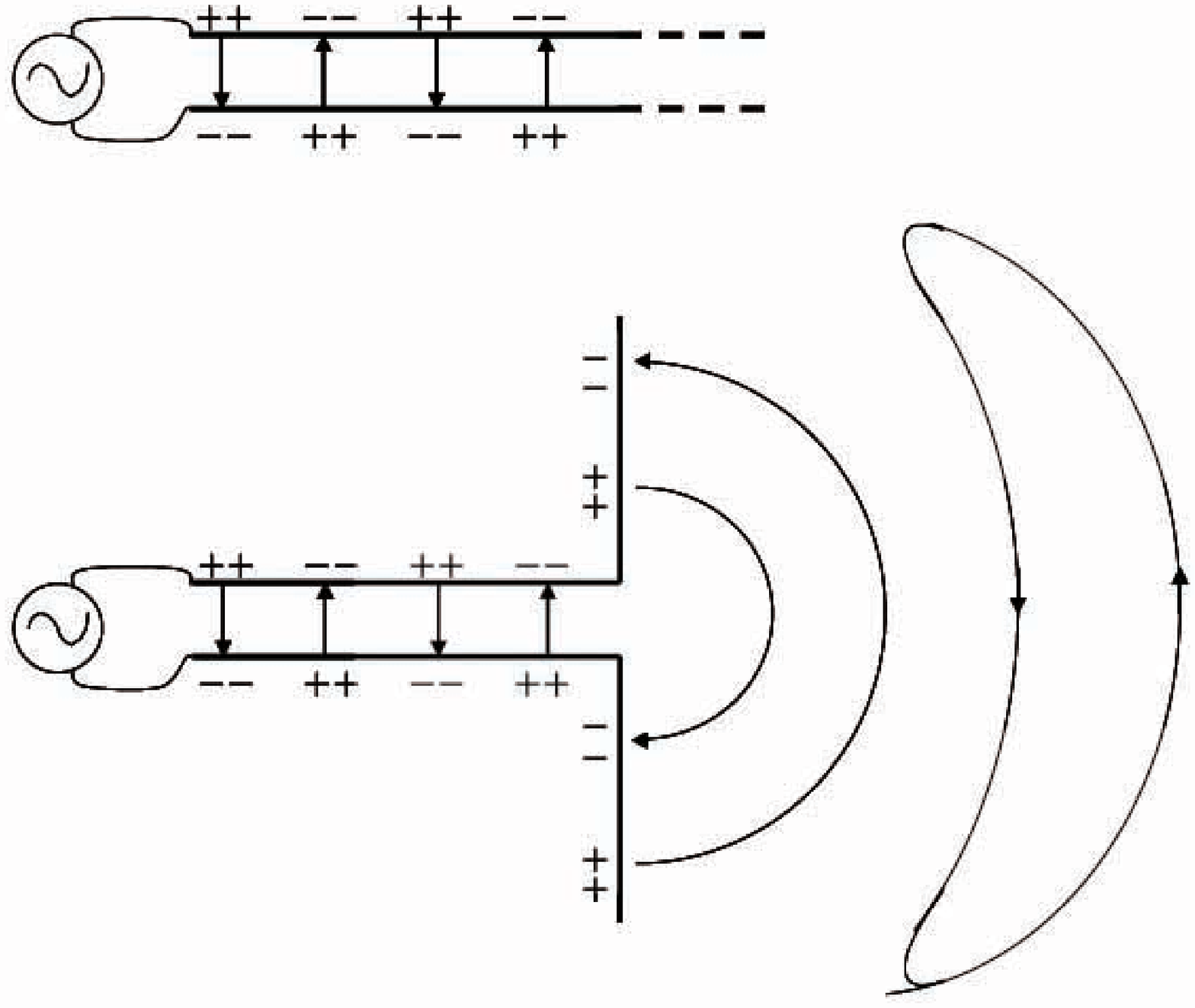}}}%
\subfigure[]{
\resizebox*{4cm}{!}{\includegraphics{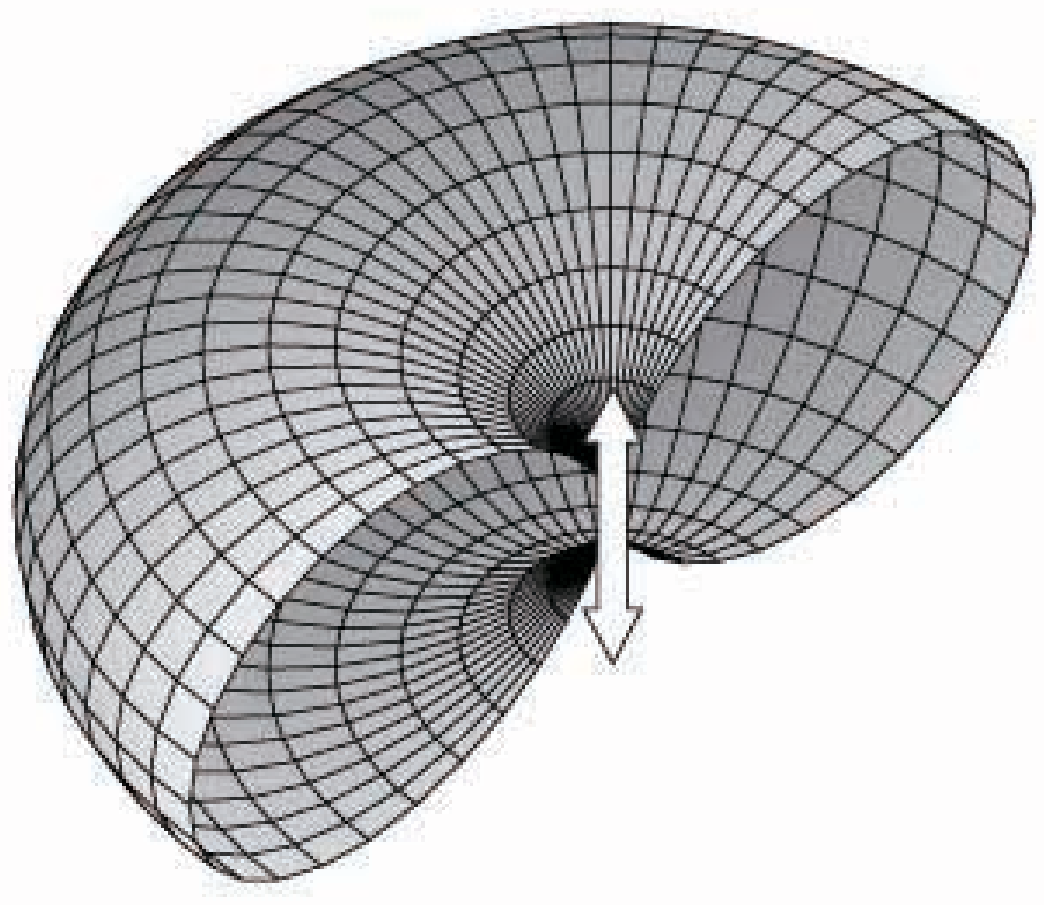}}}%
\caption{Dipole antenna: (a) two wire transmission lines.
Alternating currents in two wires have the same magnitude and phases with $180^{o}$ difference. Confined oscillating
electric field in parallel lines leaves (radiates) from lines when lines open up in opposite directions.
(b) Radiation pattern of an oscillating dipole }%

\end{minipage}
\end{center}
\end{figure}

\subsection{surface plasmon polaritons}
In finding the resonant condition of a monopole antenna, we assumed that the traveling wave current moves at the speed of
light and stated that resonance arises when the length is an integer multiple of $\lambda /4$. Can this be true for any value
of $\lambda$, especially for optical wavelengths ? Recent experiment shows that about 80 $ nm$ tall optical monopole antenna is resonant
with green light with wavelength of $ 514  nm$ \cite{Taminiau1}. Obviously, the $\lambda /4$-rule does not work. Then, what decides the
resonance condition ?

Free space Maxwell's equation is known to be invariant under the scale transformation:
\begin{equation}
\vec{r} \rightarrow \alpha \vec{r}, ~~~ t \rightarrow
\alpha t,
\end{equation}
for some constant $\alpha$. In other words, if fields $\vec{E}(\vec{r}, t)$ and $\vec{B}(\vec{r}, t)$
are given solutions of the wave equation,
\begin{eqnarray}
\nabla^2 \vec{E}(\vec{r}, t) -  {1 \over c^2 }{\partial^2 \over  \partial t^2 } \vec{E}(\vec{r}, t) &=& 0 \nonumber \\
\nabla^2 \vec{B}(\vec{r}, t) -  {1 \over c^2 }{\partial^2 \over  \partial t^2 } \vec{B}(\vec{r}, t) &=& 0
\end{eqnarray}
we have that $\vec{E}(\alpha \vec{r}, \alpha  t)$ and $\vec{B}(\alpha \vec{r},\alpha  t)$
are also solutions. If the scale invariance also works for a monopole antenna, the $\lambda /4$-resonance condition would hold
irrespective of $\lambda $.
Scale invariance, however, breaks down in the presence of a dispersive matter such as metal.
The dielectric function of a metal is frequency dependent as evidenced in the Drude model of the dielectric function,
\begin{equation}
\epsilon_{Drude} (w) = 1 - {w_p^2 \over w^2 + i \Gamma w}
\label{dielectric}
\end{equation}
where $w_p$ is the volume plasma frequency and $\Gamma $ is a damping constant. The effect of dispersive dielectric function on
the traveling wave current may be best understood in terms of surface plasmons on a flat metal surface.

\begin{figure}
\begin{center}
\begin{minipage}{100mm}
\subfigure[]{
\resizebox*{5cm}{!}{\includegraphics{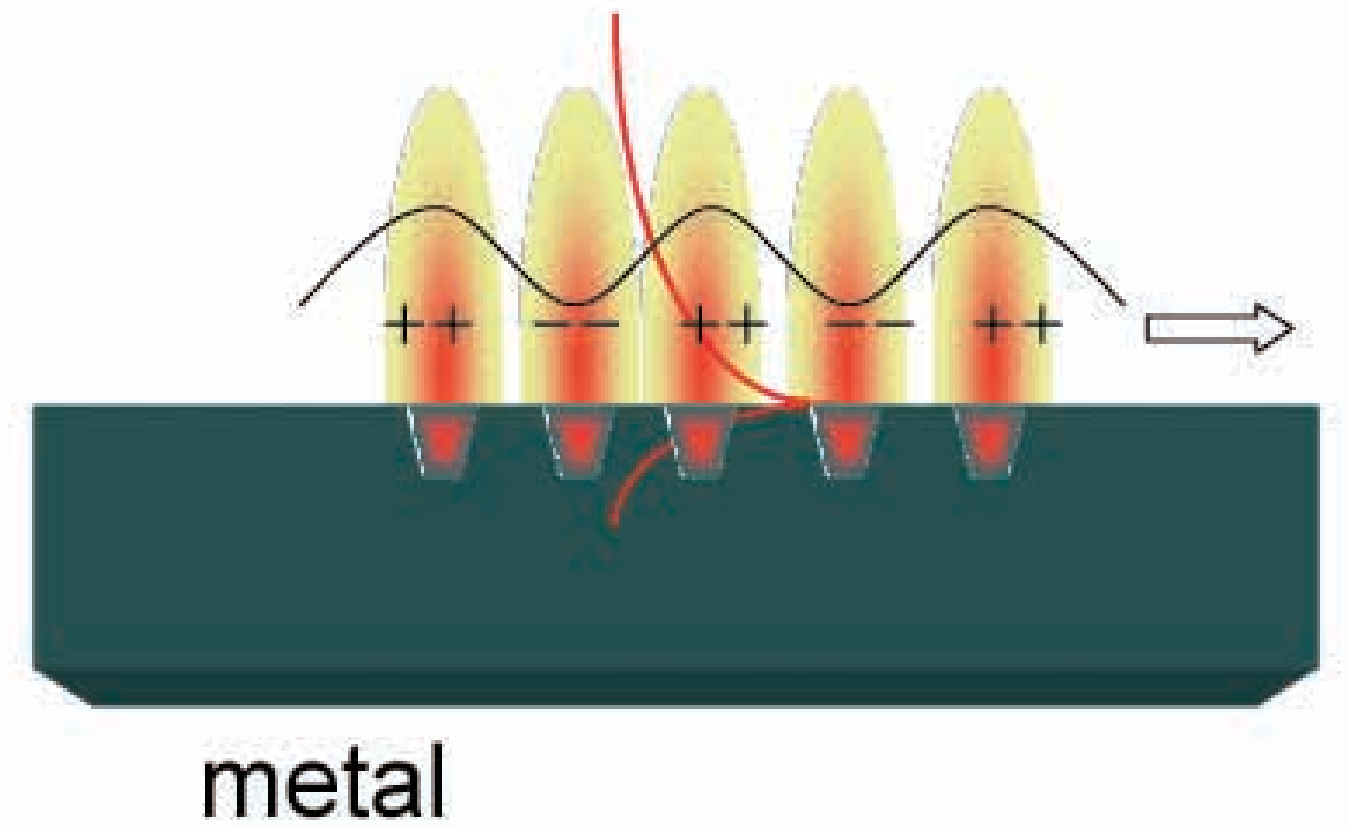}}}%
\subfigure[]{
\resizebox*{5cm}{!}{\includegraphics{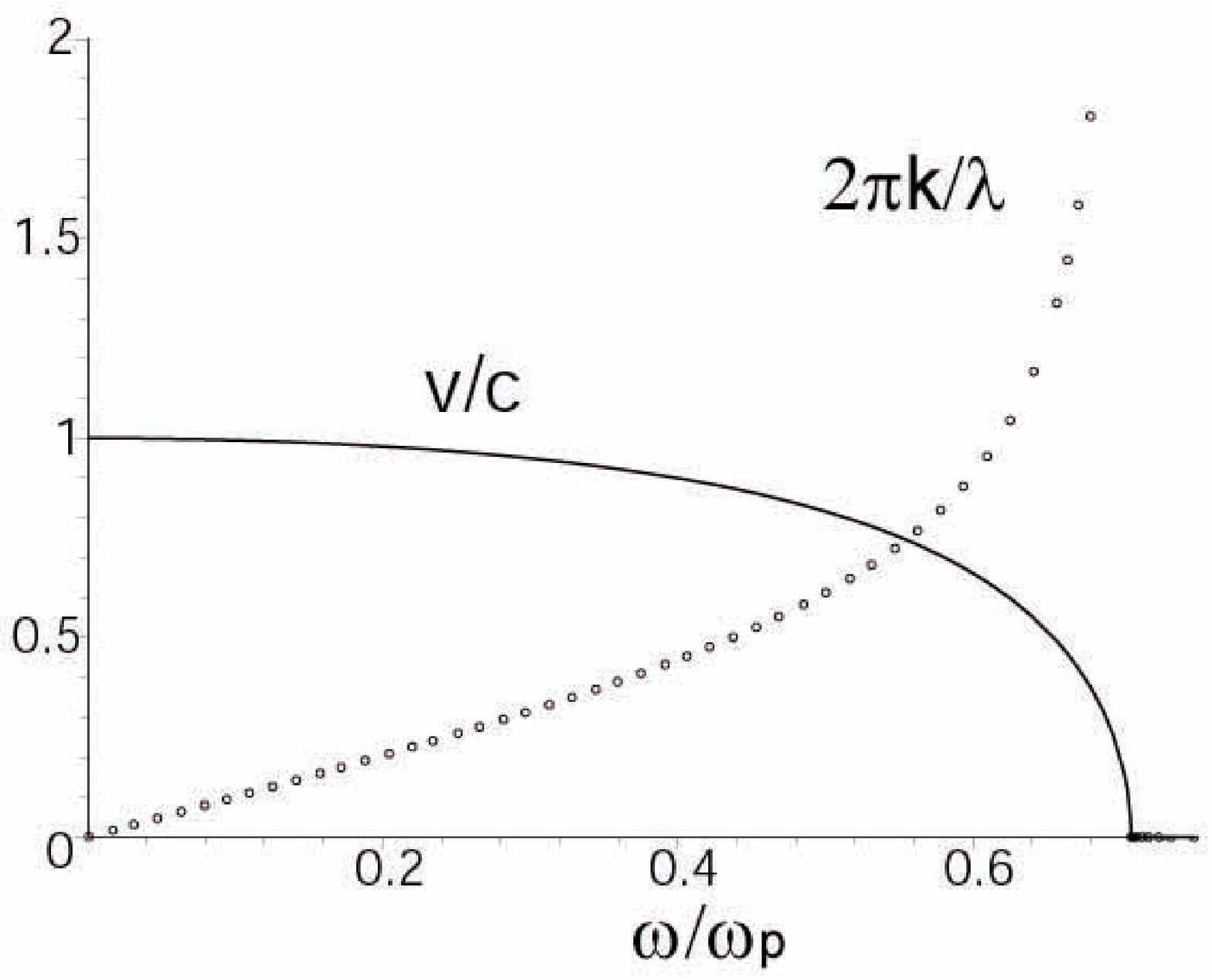}}}%
\caption{Surface plasmon polariton: %
(a) schematics of surface plasmon polariton. (b) The dispersion relation and the frequency dependent phase velocity of SPP.}%

\end{minipage}
\end{center}
\end{figure}

Surface plasmon polariton(SPP) is a surface electromagnetic wave that propagates in a direction parallel to the metal surface.
Consider the metal/dielectric interface at $ z=0$ (metal part $ z < 0$) and a surface electromagnetic wave of the form,
\begin{eqnarray}
z>0 ~~~~
&& \vec{H}= (0, ~ H_{y}, ~0)e^{i(k_{x} x + k_{z}z-wt)} \nonumber \\
&& \vec{E} = (E_{x}, ~ 0, ~E_{z})e^{i(k_{x} x + k_{z}z-wt)} \nonumber \\
z < 0 ~~~~
&& \vec{H}_{m} = (0, ~ H_{my}, ~0)e^{i(k_{mx} x -k_{mz} z-wt)} \nonumber \\
&& \vec{E}_{m} = (E_{mx}, ~ 0, ~E_{mz})e^{i(k_{mx} x -k_{mz}z-wt)} ,
\end{eqnarray}
where the subscript $m$ denotes metal. The wavevector components satisfy
\begin{eqnarray}
k_{x}^2+k_{z}^2 &=& \epsilon k_{0}^2 \nonumber \\
k_{mx}^{2}+k_{mz}^{2} &=& \epsilon_{m} k_{0}^2 .
\label{wavevector}
\end{eqnarray}
Requiring the continuity of tangential components $\vec{E}$ and $\vec{H}$ at the interface $z=0$, we have
\begin{eqnarray}
 k_{x} &=& k_{mx} \nonumber \\
~~  E_{x} &=& E_{mx} , ~~  H_{y} = H_{my}.
\label{continuity}
\end{eqnarray}
From the Maxwell's equation $ \nabla \times \vec{H} =  \epsilon \partial \vec{E}/\partial t $, we find the relation
\begin{eqnarray}
k_{z}H_{y}&=& -w \epsilon E_{x} \nonumber \\
-k_{mz}H_{my} &=& -w \epsilon_{m} E_{mx}  ,
\label{EandH}
\end{eqnarray}
which  together with equation (\ref{continuity}) leads to
\begin{equation}
{k_{z} \over \epsilon}= -{k_{mz} \over \epsilon_{m}}.
\label{kz}
\end{equation}
Combining equations (\ref{wavevector}) and (\ref{kz}), we finally obtain the dispersion relation
\begin{eqnarray}
k_{z} = i \sqrt{-\epsilon^2 \over \epsilon + \epsilon_{m}} {w \over c} , ~~~
k_{mz} = i \sqrt{ - \epsilon_m^2 \over \epsilon + \epsilon_{m}} {w \over c} , ~~~
k_{x} = \pm \sqrt{ \epsilon \epsilon_{m} \over \epsilon + \epsilon_{m}} {w \over c}.
\end{eqnarray}
where the choice of sign in taking the square root is made in the following way;
we assume for simplicity that the damping coefficient of the dielectric function in (\ref{dielectric}) is negligibly small so that
it may be neglected. We also restrict to the angular frequency $w$ smaller than the plasmon frequency $w_p$. Then, $\epsilon_m$ and
$\epsilon_m + \epsilon$ are negative and quantities inside the square root are all positive. We have chosen positive sign for
$k_{z} $ and $k_{mz}$ which presents a physically sensible solution localized near $z=0$.
As illustrated in Fig. 4a, imaginary $k_z$ and $k_{mz}$ make field decay exponentially as we move away from the interface so that field is localized
near $z=0$.
In contrast, $k_{x}$ is real. This means that field propagates along the surface with velocity $v_{sp}$ and
effective wavelength $\lambda_{sp}$,
\begin{equation}
v_{sp} = \sqrt{ \epsilon + \epsilon_{m} \over \epsilon \epsilon_{m} } c,
 ~~ \lambda_{sp} = \sqrt{ \epsilon + \epsilon_{m} \over \epsilon \epsilon_{m} } \lambda.
 \label{lambdasp}
\end{equation}
This surface bound electromagnetic wave is the well-known surface plasmon polariton. The dispersion relation and the phase
velocity of SPP are given in Fig. 4b.

The divergence free condition of electric field requires that
\begin{equation}
E_{z} = - {k_{x} \over k_{z}}E_{x} = \pm \sqrt{\epsilon_{m} \over \epsilon}E_{x},
\end{equation}
which implies that both $E_{x}, E_{z}$ are nonvanishing for a metal having finite conductivity.  This in turn implies that there exists
nonzero surface charge density $\sigma =\epsilon_{0} E_z e^{i(k_x x -wt)} $. Thus SPP accompanies collective charge
oscillations driven by light.
Note that for a highly conducting metal, $\epsilon_{m}$ is large and $k_{z}$ becomes negligibly small so that the localization is very weak.
Therefore, in the perfectly conducting limit, SPP ceases to exist if the surface is an infinite plane.
Fig. 4b shows that the velocity of SPP decreases as frequency increases.
In fact, it reaches to zero at the critical frequency $w_{sp}$,
\begin{equation}
\epsilon_{s} + \epsilon_{m}(w_{sp}) =0 ~~ \mbox{ or } ~~ w_{sp} = {w_{p} \over \sqrt{ 1+ \epsilon_{s}}} .
\label{SPresonance}
\end{equation}
The effective wavelength $\lambda_{sp}$ also shrinks as frequency increases towards $w_{sp}$.
This brings a hope that we could possibly use $\lambda_{sp}$ to explain the discrepancy between
the resonance length $L_{res}$ of a monopole antenna and the quarter wavelength ($L_{res} < \lambda /4$).
In other words, can we expect that $L_{res} = \lambda_{sp} /4$ ? To check this, we evaluate $\lambda_{sp}$
for an aluminium metal plane. The dielectric constant of aluminium at wavelength 514 nm is
$\epsilon_{m} = -38.4 + 10 i   $, obtained by fitting the Drude model to experimental
data \cite{CRC}. Then, Eq. (\ref{lambdasp}) results in $ \lambda_{sp} =508 nm$. Though this value smaller
than the free space wavelength $514 nm$, it is not sufficiently small to account for the experimentally observed
resonance length $L_{res} \approx 80 nm$ $ ( < \lambda_{sp}/4 = 127 nm)$.
One reason for this discrepancy is that we considered surface traveling waves on a flat metal surface
while the monopole antenna is made of a rod with diameter smaller than wavelength.

\subsection{Cylindrical surface waves}

\begin{figure}
\begin{center}
\begin{minipage}{100mm}
\subfigure[]{
\resizebox*{5cm}{!}{\includegraphics{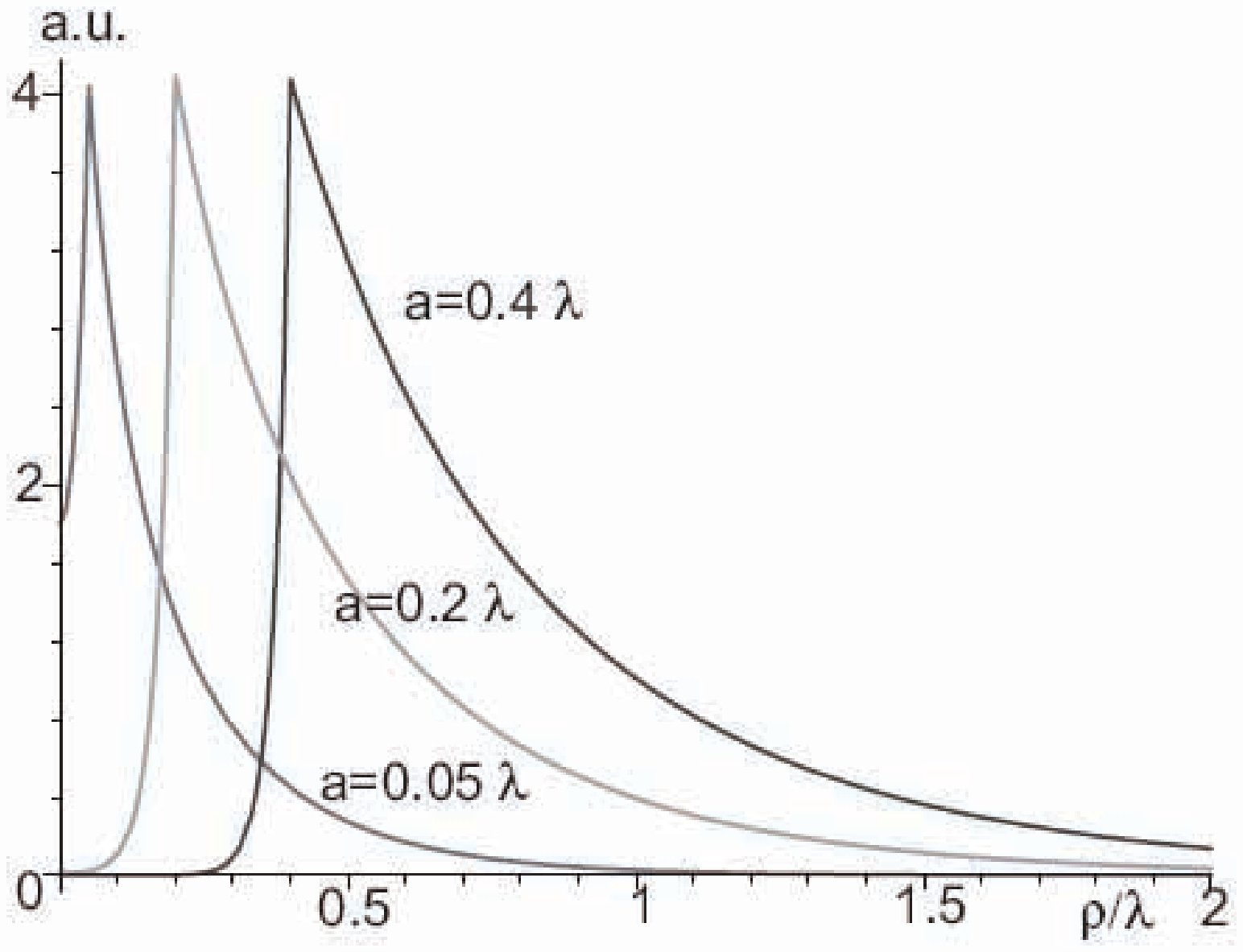}}}%
\subfigure[]{
\resizebox*{5cm}{!}{\includegraphics{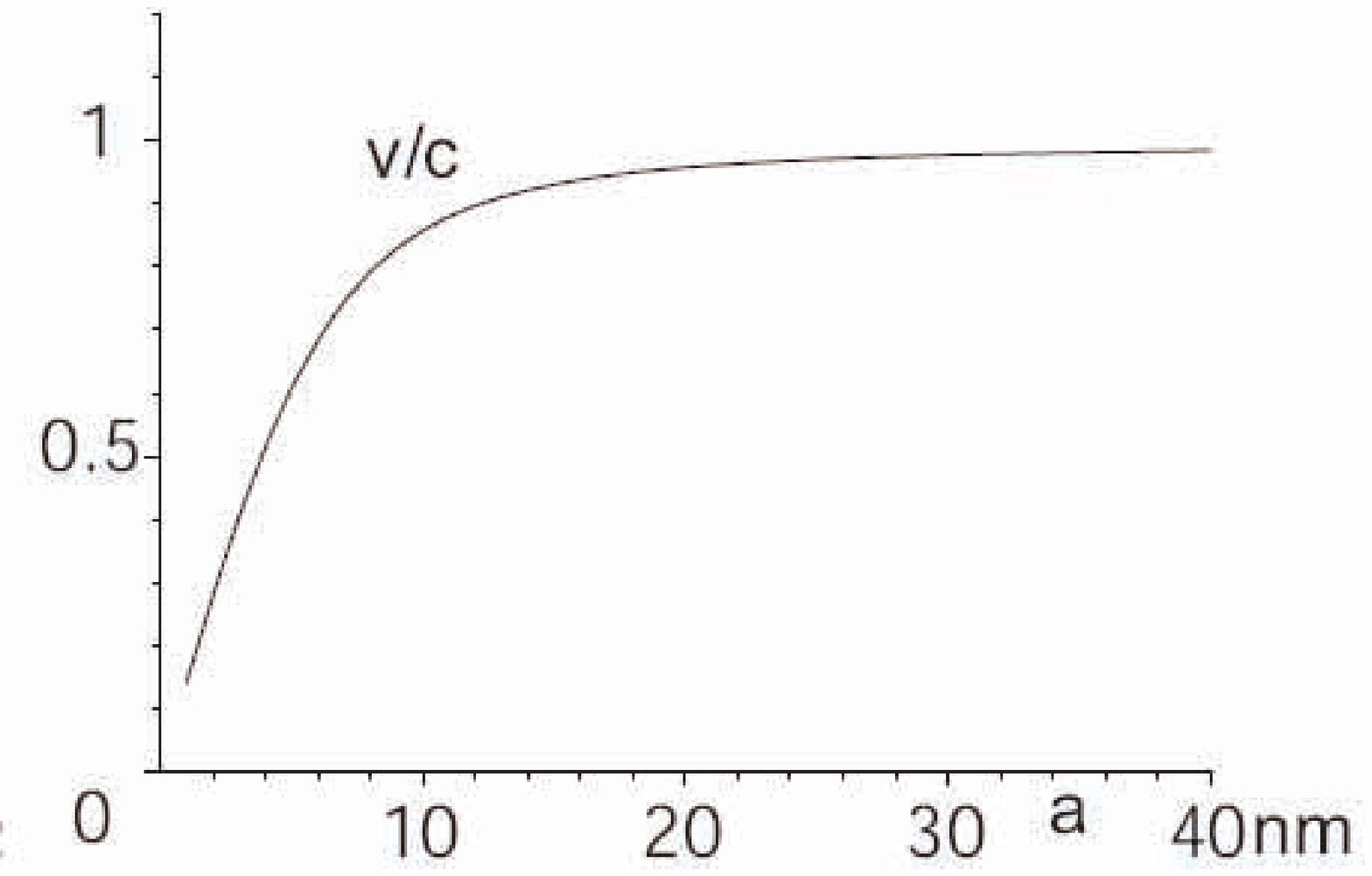}}}%
\caption{Cylindrical surface plasmon polaritons %
(a) Lowest TM modes with the radius of metal cylinder $a = 0.05 \lambda, 0.2 \lambda, 0.4 \lambda$. (b)
Phase velocity of surface waves by varying radius $a$. Smaller radius makes surface waves to stay more inside a metal
so as to reduce the phase velocity.
 }%

\end{minipage}
\end{center}
\end{figure}

In order to take into account of the shape effect, we need to find travelling waves
along the surface of a cylinder having radius $a$. In cylindrical coordinates, field components
inside and outside of a cylinder can be written as
\begin{eqnarray}
E_{z}(\rho , z) &=& \int_{-\infty }^{\infty} c_{1} J_{0}(\kappa_{m} \rho )e^{-i\alpha z} d\alpha
\mbox{ for } \rho < a \nonumber \\
&=& \int_{-\infty }^{\infty} c_{2} H_{0}^{(2)}(\kappa_{0} \rho )e^{-i\alpha z} d\alpha
\mbox{ for } \rho > a \nonumber \\
H_{\phi }(\rho , z) &=& {iw \epsilon_{m} \over \kappa_{m}} \int_{-\infty }^{\infty}
c_{1} J_{1}(\kappa_{m} \rho )e^{-i\alpha z} d\alpha \mbox{ for } \rho < a \nonumber \\
                    &=& {iw \epsilon_{0} \over \kappa_{0}} \int_{-\infty }^{\infty}
c_{2} H_{1}^{(2)}(\kappa_{0} \rho )e^{-i\alpha z} d\alpha \mbox{ for } \rho > a ,
\end{eqnarray}
where $\kappa_{m}^2 = \epsilon_{m} k_{0}^2 - \alpha^2, ~ \kappa_{0}^2 = k_{0}^2 - \alpha^2 $ and
$c_{1}, c_{2}$ are undetermined coefficients.
$J_{m}$ and $H^{(2)}_{m} $ are cylindrical Bessel functions of the first kind and cylindrical
Hankel functions of the second kind respectively.
Requiring the continuity of fields at $\rho = a$ leads to
\begin{equation}
AC\equiv \begin{pmatrix} J_{0}(\kappa_{m}a) & - H^{(2)}_{0}(\kappa_{0}a) \cr
 {iw\epsilon_{m} \over \kappa_{m}} J_{1}(\kappa_{m}a) &
- {iw\epsilon_{0} \over \kappa_{0}} H^{(2)}_{1}(\kappa_{0}a) \end{pmatrix}
\begin{pmatrix} c_{1} \cr c_{2} \end{pmatrix} =0
\end{equation}
Nonvanishing coefficients $c_{1}, c_{2}$ exist when the determinant of matrix $A$ vanishes,
\begin{equation}
det A (\alpha ) = - {iw\epsilon_{0} \over \kappa_{0}} H^{(2)}_{1}(\kappa_{0}a) J_{0}(\kappa_{m}a)
+  {iw\epsilon_{m} \over \kappa_{m}} J_{1}(\kappa_{m}a) H^{(2)}_{0}(\kappa_{0}a) =0.
\end{equation}
The zeros of $det A (\alpha )$ represent the symmetric TM modes of the lossy dielectric cylinder.
If we denote the root corresponding to the principle mode $TM_{0}$ by $\alpha_{0}$, wavenumber $\alpha_{0}$
determines the phase velocity of a surface wave by
\begin{equation}
v_{p} = {k_{0} \over \mbox{Re} (\alpha_{0})}c.
\end{equation}
Once again, this exhibits a plasmonic effect through a surface wave propagating along a cylinder with reduced
phase velocity.
Fig. 5 shows TM mode profiles and the phase velocity of the principle mode as a function of radius.
It shows that the phase velocity and subsequently the effective wavelength decreases as the radius becomes smaller.
When $a=20 nm$,  $\lambda_{eff} = 491 nm$. Though shorter than the flat space surface plasmon wavelength,
the quarter of $\lambda_{eff}$ is still bigger than the resonance length $L_{res} \approx 80 nm$.
The actual monopole antenna, however, is not an infinite cylinder but has a rod end.
It was pointed that the reactance of the rod end would increase the antenna length effectively \cite{Novotny} and this
modifies the effective wavelength approximately by
\begin{equation}
\lambda_{eff} ={k_{0} \over \mbox{Re} (\alpha_{0})} \lambda - 4 a .
\end{equation}
One might think as well that the reflection of a surface wave at the rod end suffers a certain amount of phase shift
which would increase the antenna length effectively.
There could be also experimental factors contributing to the discrepancy for the measured resonance length such as
non perfectly conducting ground which however are not our concern. Instead of pursuing a specific monopole
antenna problem more(for further details, see for example \cite{Hanson}), we consider the effect of finite size
on plasmonic resonances in another well known system - metal nanoparticles.

\subsection{Local plasmon resonance and nanoparticles}
When small metallic particles are irradiated by light, the conduction electrons oscillate collectively with
oscillation frequencies determined by the density and the effective mass of electrons as well as the shape and
the size of a metal particle. The density $n$ and the effective mass $m_{e}$ of electrons determine
the macroscopic dielectric function in Eq. (\ref{dielectric}) through the bulk plasma frequency
$w_{p}=\sqrt{ne^2 /m_{e}\epsilon_{0}}$.
To relate the shape and the material property to the oscillation frequencies, we consider a simple case of
a small spherical metal particle for which the quasi-static approximation is available.
If the radius $a$ is much smaller than the wavelength $\lambda$, all electrons in a sphere can be assumed to
respond simultaneously to electric field of incoming light. In a short time scale, compared to the
time period of light, electric field remains to be constant and the interaction can be treated as an
electrostatic problem of solving the Laplace equation. Denoting the constant electric field
by $\vec{E} = E_{0}\hat{z}$, the electric field outside the sphere can be readily calculated by solving the
potential problem. Take the potential $\varphi $ of the form,
\begin{eqnarray}
\varphi = & A r \cos \phi & \mbox{ for } r < a \nonumber \\
\varphi = & (-E_0 r + B r^{-2} )\cos \phi & \mbox{ for } r > a .
\end{eqnarray}
The continuity of the tangential electric fields, $\partial \varphi /\partial \theta $,
and the normal components of the electric displacements, $\epsilon \partial \varphi /\partial r $,
at the surface of a sphere lead to
\begin{equation}
A=-{3\epsilon_{s} E_{0} \over \epsilon_{m} + 2\epsilon_{s}},
 ~~ B = {\epsilon_{m} - \epsilon_{s} \over \epsilon_{m} + 2\epsilon_{s}}a^3 E_{0}.
\end{equation}
Then, the electric field outside the sphere is written as
\begin{equation}
\vec{E}_{out} = E_{0} \hat{z} -{\epsilon_{m} - \epsilon_{s} \over \epsilon_{m} + 2\epsilon_{s}}a^3 E_{0}
\Big({\hat{z} \over r^3} - {3 z \vec{r} \over r^5}\Big) ,
\label{dipole}
\end{equation}
where $\epsilon_{m}$ and $\epsilon_{s}$ denote dielectric constants of metal and surrounding material respectively.
Apart from the incident field, this is the electric field of an induced dipole with the particle polarizability
\begin{equation}
\alpha = 4\pi a^3 {\epsilon_{m} - \epsilon_{s} \over \epsilon_{m} + 2\epsilon_{s}}.
\end{equation}
At first sight, this quasi-static approximation looks to be really static and thus it can not explain
the resonance frequency which is a dynamical quantity. The dielectric constant $\epsilon_{m}$ of metal
is, however,  strongly dependent on frequency. A more rigorous solution of Maxwell equation without
the electrostatic approximation shows that the dipole field in (\ref{dipole}) is a radiating dipole and the
scattering efficiency is proportional to $|(\epsilon_{m} - \epsilon_{s}) / (\epsilon_{m} + 2 \epsilon_{s})|^2$.
The plasmon resonance in scattering occurs when the real part of the denominator
vanishes such that
\begin{equation}
Re(\epsilon_{m} + 2\epsilon_{s})=0.
\end{equation}
Or, if we neglect the damping coefficient $\Gamma$, when the dipole plasmon resonance frequency $w_{dp}$ is
\begin{equation}
w_{dp} ={ w_{p}\over \sqrt{1+2\epsilon_{s}}}.
\label{dpResonance}
\end{equation}
Comparison of $w_{dp}$ with the surface plasmon resonance frequency $w_{sp}$ in (\ref{SPresonance})
shows the dependence of ersonance frequency on the shape. A similar dependence can be found for more general shapes.
For an ellipsoidal metal particle with principal axes $a, b, $ and $c$, the particle polarizability $\alpha $ can be found
in terms of geometrical depolarization factors $L_{i}$ along the principal axes \cite{Kreibig, Bohren},
\begin{equation}
\alpha = {4 \over 3}\pi abc  {\epsilon_{m} - \epsilon_{s} \over \epsilon_{s} + L_{i}(\epsilon_{m}-\epsilon_{s})},
 ~ \sum_{i=1}^{3} L_{i} = 1
\end{equation}
and the dipole plasmon resonance frequency becomes
\begin{equation}
w_{dp} = { w_{p}\over \sqrt{1+\epsilon_{s}(1-L_{i})/ L_{i}}}.
\label{ellip}
\end{equation}
This reduces to (\ref{dpResonance}) for a spherical case for which $L_{1} = L_{2} = L_{3} =1/3$.
Note that the dipole plasmon resonance frequency in (\ref{dpResonance}) does not depend on the particle size $a$.
This is due to the quasi-static approximation which is valid only for particles much smaller than wavelength.
For larger particles, the dipole plasmon resonance frequency redshifts due to retardation effects \cite{MMeier}
and radiation damping effects \cite{Wokaun}. Retardation effects arise since the conduction electrons in a large particle
do not move simultaneously in phase and this leads to a reduction of depolarization field. Radiation damping contributes to the
plasmon damping resulting the broadenng of resonance spectrum. Retardation and radiation damping effects
appear as the lowest-order corrections to the quasi-static approximation. In particular, the dipole plasmon resonance
frequency is corrected by
\begin{equation}
w_{dp} \approx  { w_{p}\over \sqrt{1+2\epsilon_{s}+ 12 \epsilon_{s}^2 \pi^2 a^2 / 5\lambda^2 }}.
\end{equation}
In addition to the dipole,  for larger particles there are contributions from higher-order modes (quadrupole, octopole, etc.)
and the spectral response is also modified by the higher-order effects. All these corrections can be calculated from the Mie
theory of scattering \cite{Mie}.

\section{Field enhancement and optical aperture antennas}
Most antennas were of wire type (monopoles, dipoles, helices etc.) before World War II. During the War, new antenna technology
was launched and aperture type antennas(slits, slots, waveguides, horns, reflectors) operating at microwave frequencies were introduced.
Aperture type antennas are particularly important in optical applications. Optical bowtie slot antenna is one good example of aperture
type antenna that has a better control of the radiation pattern compared to wire antennas.
In fact, the uprise of plasmonics has begun with the measurement of extraordinary transmission of light through periodic arrays of holes
in an opaque metal film \cite{Ebbesen}. The role of surface plasmon polaritons in periodic arrays of holes or slits has
received much attention \cite{Porto,Treacy,Cao,Lee1}.
After all, periodic array of holes are nothing but a phased-array optical slot antenna.

\subsection{Circular and rectangular holes}
How does light pass through a hole of size smaller than or comparable to wavelength? This is a well known question in diffraction theory
treated in optics texts books. Fraunhofer diffraction provides far field configurations by utilizing the Huygens-Fresnel principle
stating that light splits into outgoing spherical waves when passed through a hole or a slit.
When a hole is perforated in a metal, other important contribution arises. Impinging light drives surface current and accumulates charge
at the edges of the hole. Accumulated charge in turn enhances electric field inside the hole and subsequently the far field diffracted light.
Therefore a metallic aperture can focus and transmit electromagnetic waves with increased efficiency. In other words, it acts as an antenna.
This problem was first considered by Bethe in the advancement of microwave technology during the World War II.
Bethe developed his theory of light diffraction by idealizing the structure as the circular aperture in a perfectly conducting screen
of zero thickness. The transmission efficiency, defiend as transmission cross section divided by hole area, was obtained by Bethe
in the long wavelength limit and corrected later to include higher-order terms by Bouwkamp \cite{Bouwkamp}. It is given by
\begin{equation}
{\sigma \over \pi a^2 } = {64 (ka)^4 \over 27 \pi^2 }\Big[ 1 + {22 \over 25}(ka)^2 + \cdots \Big]
\end{equation}
where $k =2\pi /\lambda $ is the wavenumber of incoming wave and $a$ is the radius of a hole. It shows that the transmission
efficiency scales as $(a/\lambda )^4$ for $a < \lambda$ and drops quickly as $a$ becomes small. In the optical application of
aperture antennas, focusing light and increasing transmission are important issues while
in the microwave application, radiation pattern and directivity of waves are important issues.
According to Bouwkamp, the near field $z$-component of electric field at the shadow side ($z=0$), or the surface charge per permittivity,
is written as
\begin{equation}
E_{z} = { \sigma \over \epsilon_{0}} = E_{0}{4i ka \over 3}{a/\rho \over \sqrt{\rho^2 /a^2 -1}}\cos \phi ,
\end{equation}
where $\rho (>a )$ is the radial coordinate, $\phi $ is the azimuthal angle, and $E_{0}$ is the amplitude of incoming electric field.
The angular $\cos \phi $-dependence is a characteristic of a dipole emission pattern.
This is verified both experimentally and numerically using the Finite Difference Time Domain(FDTD) method as shown in Fig. 6.
Indeed, Babinet's principle relates an aperture antenna with a dipole antenna of wire type.
Babinet's principle states that when a field behind a perfectly conducting screen with an aperture is added to the field of
a complementary structure which is obtained by replacing aperture by screen and screen by aperture,
then the sum is equal to the field where there is no screen. More rigorously, the (vectorial) Babinet's principle states that the
diffracted electromagnetic fields caused by a screen with apertures and the source field $(E^{(0)}, B^{(0)})$ are identical to those of
a complementary structure excited by the source field with opposite polarization characteristics
$(E^{(0)}_{c}=cB^{(0)}, B^{(0)}_{c} = - E^{(0)}/c)$. In particular, this implies that the radiation pattern from a circular hole
is the same as that of a circular disk with its driving electric field having an opposite polarization.
Since circular disk exhibits resonant behavior, we expect that resonance also arises as the size of hole $a$ becomes comparable to
wavelength $\lambda$. In order to understand the resonance in aperture antennas, we will consider a simpler case of
slot antenna, having a rectangular shape aperture.

\begin{figure}
\begin{center}
\begin{minipage}{100mm}
\subfigure[]{
\resizebox*{5cm}{!}{\includegraphics{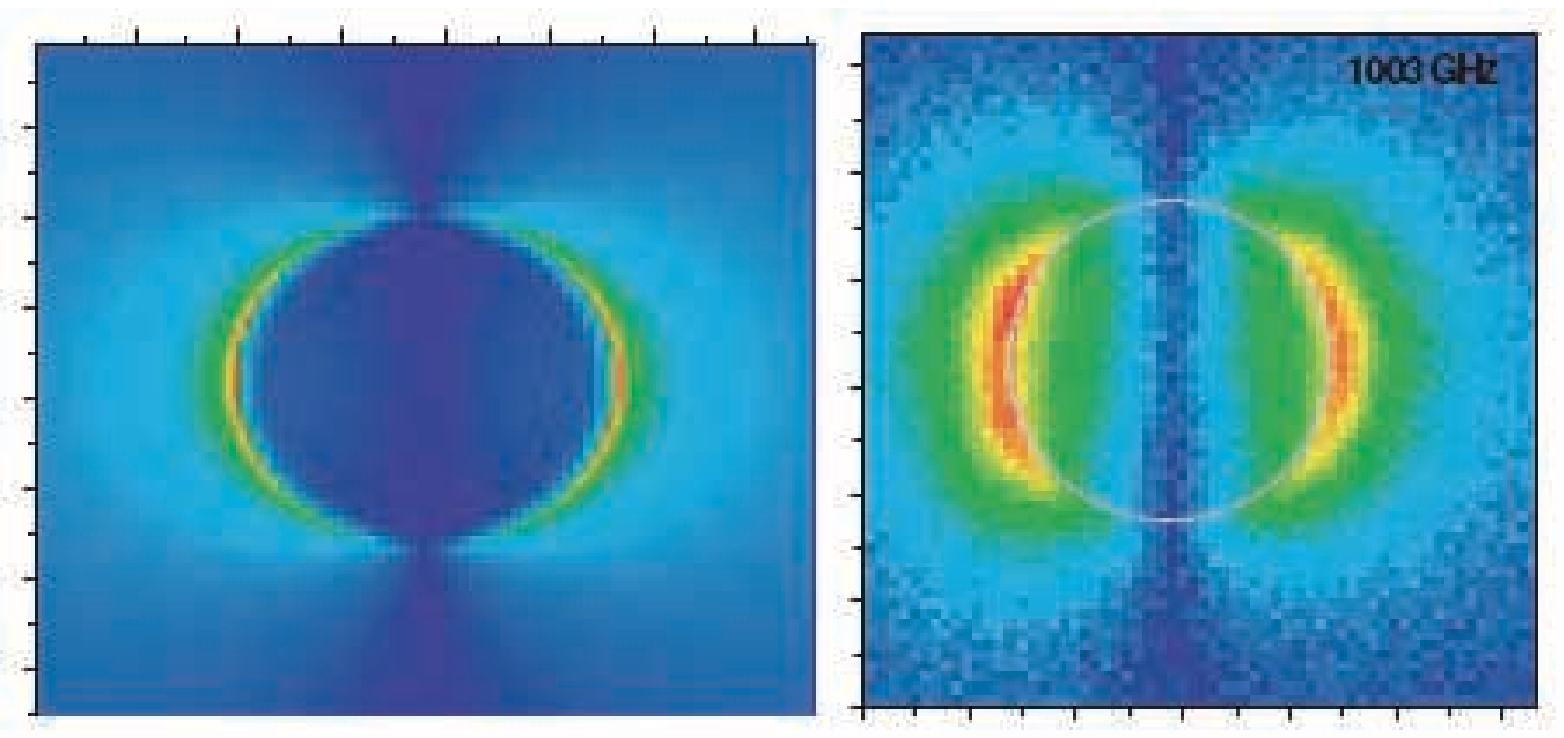}}}%
\subfigure[]{
\resizebox*{5cm}{!}{\includegraphics{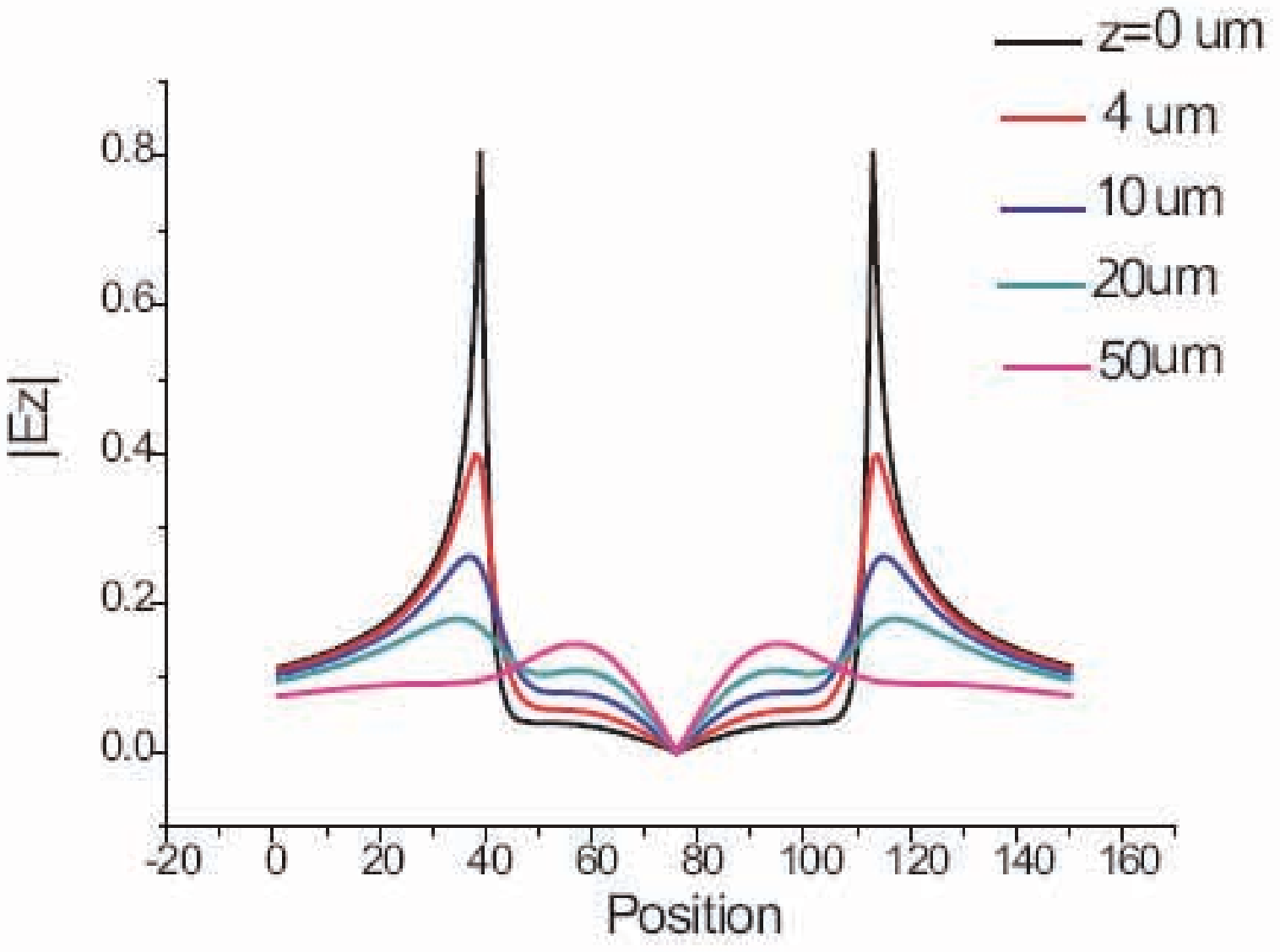}}}%
\caption{Configuration of electric field component $E_{z}$(at $z=0$) normal to the 150 $\mu m$ diameter hole:
(a) FDTD calculation (left) and experimental result using THz measurement (right).
Electro-optic sampling technique used in the experiment measures field component slightly away from $z=0$.
(b) Configurations of electric field at various values of $z$. }%

\end{minipage}
\end{center}
\end{figure}

Consider a TM polarized plane wave impinging normally on a perfectly conducting screen with a rectangular hole.
The size of a rectangular hole is $a \times b$, the thickness of screen is $h$, and the amplitude of incoming
electric field is $E_{0}$.
The lowest TE$_{01}$ mode of electric field inside the hole, directed along the $y$-axis, has the form
\begin{equation}
E_{x} =0, ~ E_{z} = 0, ~  E_{y} =  \sin\Big({ \pi x \over a}\Big)
\Big[ -E_L \cos (\mu z)  + E_R \sin (\mu z) \Big].
\end{equation}
For simplicity, we make the single mode approximation such that the electric field inside the hole is governed by the lowest TE$_{01}$
mode only.
Matching tangential components of electric and magnetic fields at the opening $(z=0)$ and the ending $(z=h)$ interfaces of hole
and also imposing the boundary condition of vanishing tangential electric field on a perfect conductor,
one can readily determine the coefficients $E_L $ and $E_R$ so that
\begin{eqnarray}
E_y = {8  k E_{0} \over  \pi } \sin \Big( {\pi x\over a}\Big) {i\mu \cos(\mu z) + k W \sin(\mu z )
\over  (\mu^2 + k^2 W^2  ) \sin(\mu h) + 2i \mu  k W \cos(\mu h) }, ~~ 0 < z < h
\label{Efield}
\end{eqnarray}
where $k=2\pi / \lambda, ~ \mu =\sqrt{k^2-(\pi /a)^2}$. $W$ representing the coupling between incoming wave and the TE$_{01}$ mode
is given by
\begin{eqnarray}
W &=&  \int dk_x dk_y {k_y^2 + k_z^2 \over k k_z }
{2\over ab}\Big[ \int_{0}^{a}dx \int_{0}^{b}dy \sin\Big({ \pi x \over a}\Big)e^{i\Theta }\Big]^2 \nonumber \\
\Theta &=& k_x (x-a/2) + k_y (y-b/2), ~~ k_z^2 = k^2 -k_x^2 -k_y^2,
\end{eqnarray}
When the hole is of subwavelength scale ($a, b < \lambda $), $W$ is approximately equal to $ 32 ab k^2 / 3\pi $.
For the screen of nearly zero thickness ($h \ll \lambda $), the electric field at the center of the hole $(x=a/2, z=0)$ is approximately
\begin{equation}
E_y \approx {8 ikE_{0}/\pi \over h[k^2 - (\pi /a)^2 + k^2 W^2 ] + 2ik W}.
\label{Ecenter}
\end{equation}
Now, we are ready to understand the resonance nature of a rectangular aperture. If we assume that the slot is thin ($ b \ll a < \lambda $)
so that $W$ is small, Eq. (\ref{Ecenter}) shows that resonance occurs when $ k \approx \pi /a$ or the long side of slot is half wavelength long
$a \approx \lambda /2$. Note that this is the resonance condition of a dipole antenna of wire type. The direction of electric field is
along the $y$-axis which is perpendicular to the long axis of slot. This may be compared with the wire type antenna where the electric field
is along the direction of wire. Thus we assured the Babinet's principle in our case which is illustrated in
Fig. 7a. We observe that near resonance electric field is approximately equal to
\begin{equation}
 E_y \approx {3 E_0 \over 8 ab k^2 },
\end{equation}
which shows that electric field is inversely proportional to the slot width $b$. Consequently, electric field
multiplied by the area of slot remains constant and slot antennas focus electromagnetic field.
This focusing property of slot antenna was predicted theoretically \cite{Garcia-Vidal} and confirmed experimentally using
terahertz waves \cite{DSKim}. In Fig. 7b, experimentally measured images of $E_{y}$ are presented showing
resonantly excited fields and the reciprocal dependence on slot width at resonance.
When moved away from resonance, the magnitude of enhancement decreases. Nonetheless, enhancement plays an important
role as can be seen in the case of an infinitely long slot, that is, slit.

\begin{figure}
\begin{center}
\begin{minipage}{100mm}
\subfigure[]{
\resizebox*{5cm}{!}{\includegraphics{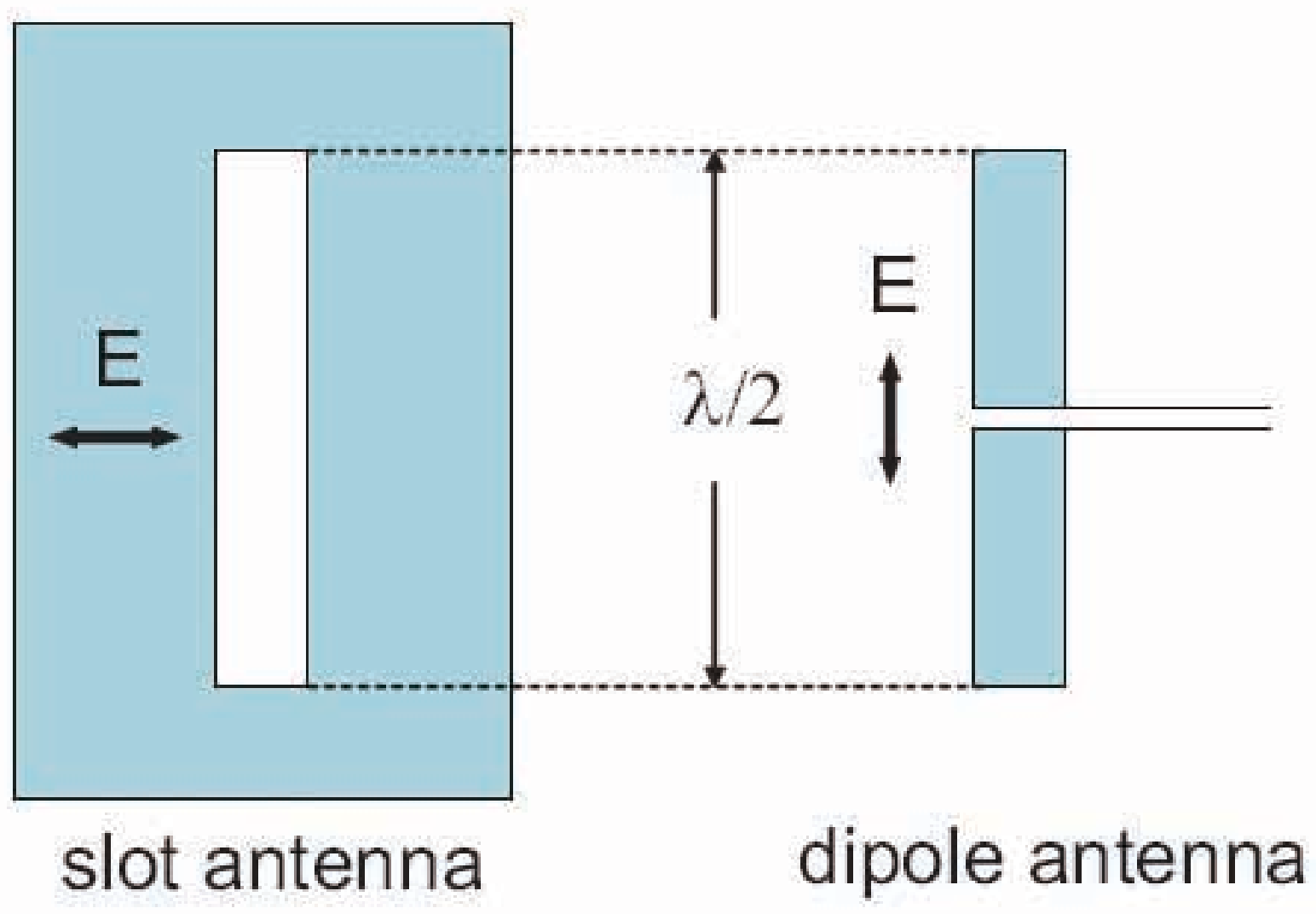}}}%
\subfigure[]{
\resizebox*{5cm}{!}{\includegraphics{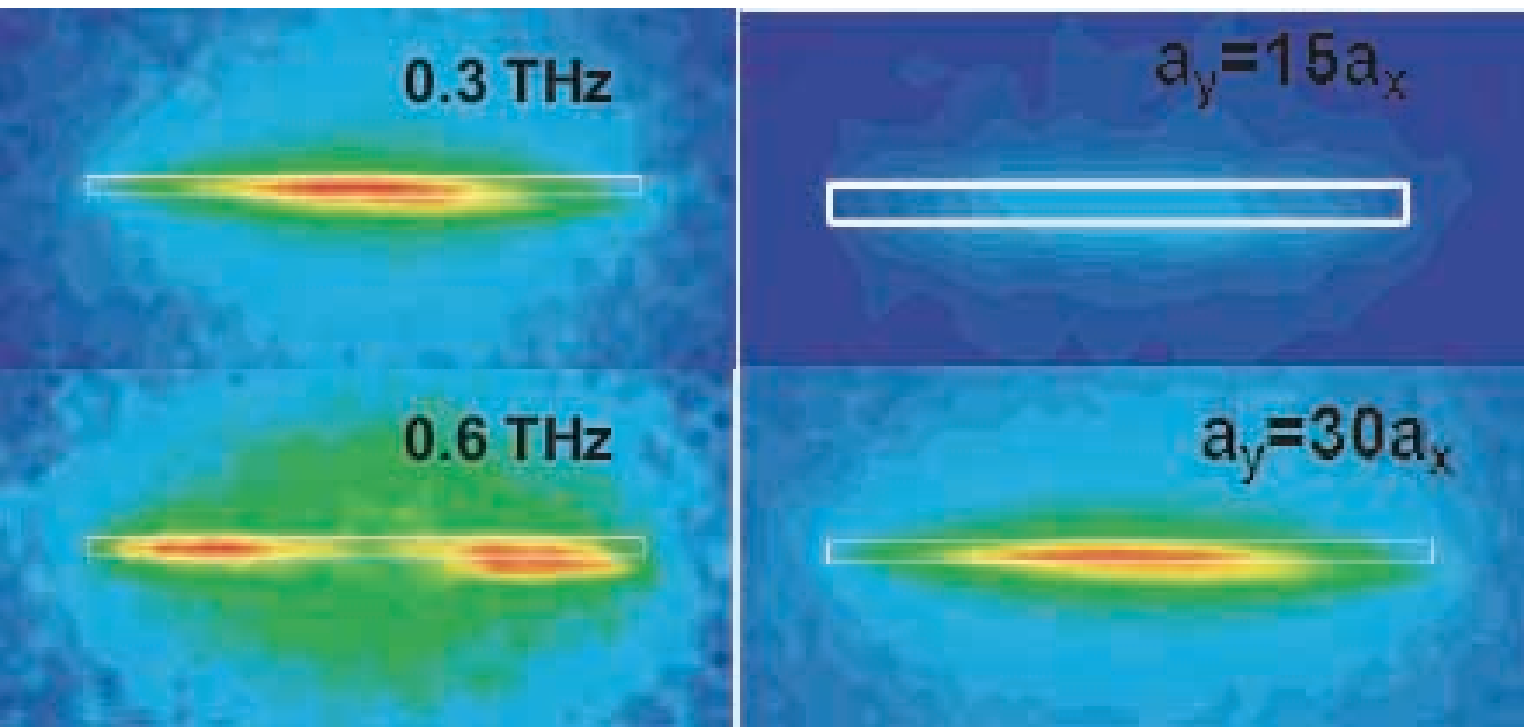}}}%
\caption{(a) Slot antenna and its Babinet complementary dipole antenna. (b) Images of $E_{y}$ component:  resonantly excited electric field (left) at a rectangular
aperture of size $10 \mu m \times 300 \mu m$ with frequencies $f=0.3 $ THz and $ 0.6 $ THz and (right) at
apertures with $a_y = 300 \mu m$ and $ f= 0.3 $ THz.  }%

\end{minipage}
\end{center}
\end{figure}

\subsection{Field enhancement without resonance: single slit case}
The expression of electric field in a slit may be obtained from Eq. (\ref{Efield}) by taking the limit $ a\rightarrow \infty$
and replacing the TE$_{01}$ mode by a TEM mode which can be done by changing $\sin(\pi x/a)$ into a constant.
We may fix the constant through matching boundaries as before. The result is written as
\begin{equation}
E_y = -2iE_{0} {\cos(k z) + G \sin(k z ) \over (1-G^2 )\sin (k h) - 2 G\cos(k h) }
\end{equation}
where the coupling constant $G$ is given by
\begin{equation}
G = \int_{0}^{b}dx \int_{0}^{b}dy {k \over b}  \int ds { e^{-isx+isy}\over i\sqrt{k^2-s^2}}
\approx  bk (2 \mbox{ln}(b k /2)-3 -i \pi +2\gamma )
\end{equation}
where $\gamma=0.5772...$ is the Euler constant. When the thickness of screen $h$ is smaller than wavelength
so that $kh = 2\pi h /\lambda \ll 1$, we may approximate $E_y$ by
\begin{equation}
E_y \approx {-2iE_{0} \over (1-G^2 ) k h - 2 G },
\end{equation}
and additionally if $h$ is smaller than the slit width $b$ ($ h < b \ll \lambda $), $E_y$ can be further approximated to
\begin{equation}
E_y \approx {iE_{0} \over G } \approx {iE_{0} \over bk (2 \mbox{ln}(b k /2)-3 -i \pi +2\gamma ) } .
\label{InverseSlit}
\end{equation}

\begin{figure}
\begin{center}
\begin{minipage}{100mm}
\subfigure[]{
\resizebox*{5cm}{!}{\includegraphics{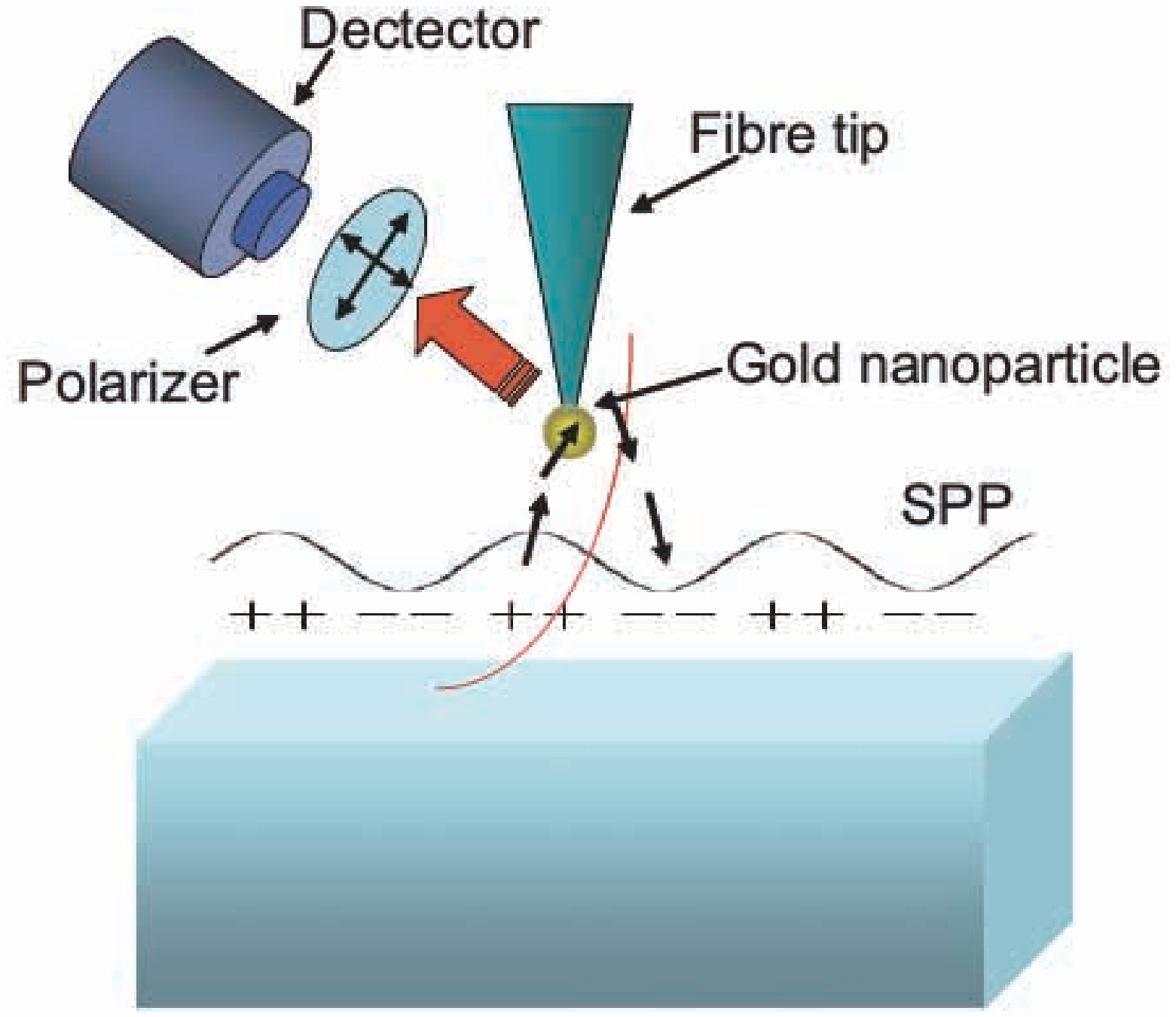}}}%
\subfigure[]{
\resizebox*{5cm}{!}{\includegraphics{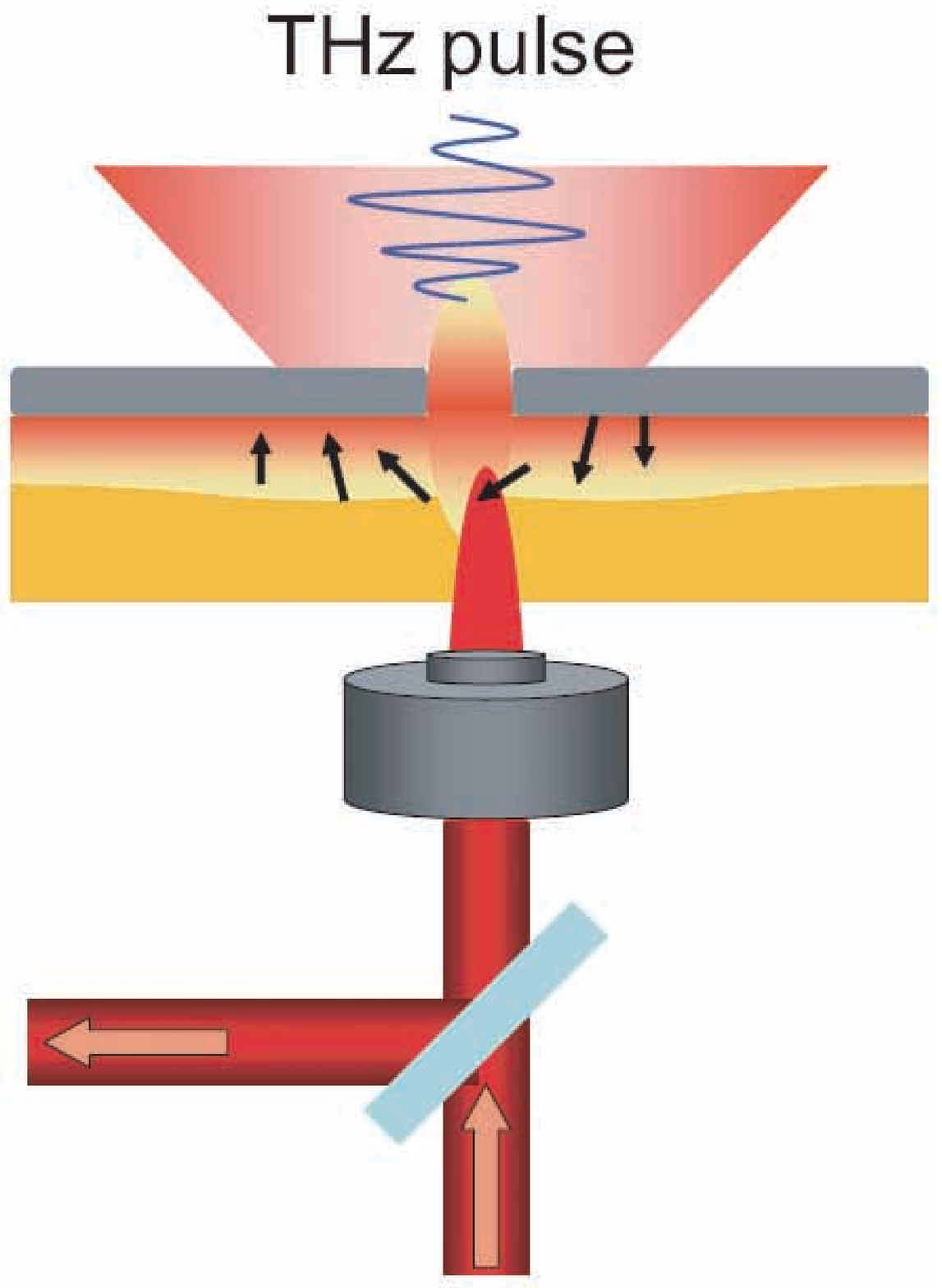}}}%
\caption{Vector field mapping: schematic of (a) Vector NSOM and (b) terahertz time-domain
spectroscopy using an electro-optic sampling technique. }%

\end{minipage}
\end{center}
\end{figure}

Once again, this shows that $E_y$ is inversely proportional to $b$ and the strong focusing of light is possible
even with a non-resonant slit system. Therefore slit also acts as an optical antenna. To understand the
physical mechanism of field enhancement, we consider the case where $h$ bigger than the slit width $b$ ($ b < h \ll \lambda $) so that
$E_y$ is approximately
\begin{equation}
E_y \approx -{2iE_{0} \over k h  }= {E_{0}\lambda \over i\pi h  }.
\label{GapE}
\end{equation}
The meaning of this expression can be understood as follows; recall that an electromagnetic wave reflected on a metal surface
induces surface current. On a perfect conductor, the induced surface current $K$ is determined by the tangential magnetic field,
\begin{equation}
\vec{K} = \vec{n} \times \vec{H}_{||}
\end{equation}
where $\vec{n}$ is the unit vector normal to the surface and  $\vec{H}_{||} $ is the tangential magnetic field just
outside the surface. The boundary condition of vanishing tangential electric field requires that $\vec{H}_{||} = 2 H_{0} $
for a normally incident plane wave with magnetic field $H_{0}=\sqrt{\epsilon_0 / \mu_0}E_{0}$. In the presence of a slit,
surface current accumulates charge at the edges of a slit as slit gap blocks the flow of surface current.
The induced charge $Q_{Ind}$ by the surface current can be readily obtained from the charge conservation law:
\begin{equation}
{\partial \over \partial t}\sigma + \nabla \cdot \vec{K} =0,
\end{equation}
where $\sigma$ is the surface charge density and $\nabla \cdot $ denotes the two-dimensional divergence. Making integration,
we find that the induced charge per unit length can be written as
\begin{equation}
Q_{Ind}  = {1\over iw} \int \vec{K}\cdot \hat{n}dl ={2\over iw} \sqrt{\epsilon_0 \over \mu_0}E_{0}
\end{equation}
where $\hat{n}$ is a normal vector to the boundary and the integration is over the unit length of a
boundary parallel to slit. We assumed the harmonic time dependence $e^{-iwt}$ for surface current $K$
and charge density $\sigma$.
For a narrow gap ($b <  h \ll \lambda  $), we expect that induced charge resides mostly on the gap surface
with surface charge density $\sigma_G = Q_{Ind}/h$ thereby forming
a parallel plate capacitor. Then the resulting electric field,
\begin{equation}
 E_y = \sigma_G / \epsilon_0  =  E_{0} \lambda / i \pi h
\end{equation}
agrees with that in Eq. (\ref{GapE}). This shows that the field enhancement inside a narrow gap indeed results from a capacitative
charging of metal edges. We learn from Eq. (\ref{InverseSlit}) that for a thin slit ( $ h < b \ll \lambda $), the field enhancement is
inversely proportional to the gap size $b$ in contrast to a thick slit where it is independent of the gap size. This is rather
surprising since it suggest that we can achieve a field enhancement without invoking resonance.
Recently, we have demonstrated that, using terahertz waves and a slit of size $\lambda /30000$,
the enhancement factor can reach to the value of 800 \cite{Seo2}.
The nonresonant enhancement may be attributed to a combined effect of the capacitative charging
and the lightning rod effect. A more rigorous account of nonresonant enhancement can be obtained by adopting the ``$\lambda-$zone"
formalism \cite{QPark}.

\begin{figure}
\begin{center}
\begin{minipage}{100mm}
\subfigure[]{
\resizebox*{5cm}{!}{\includegraphics{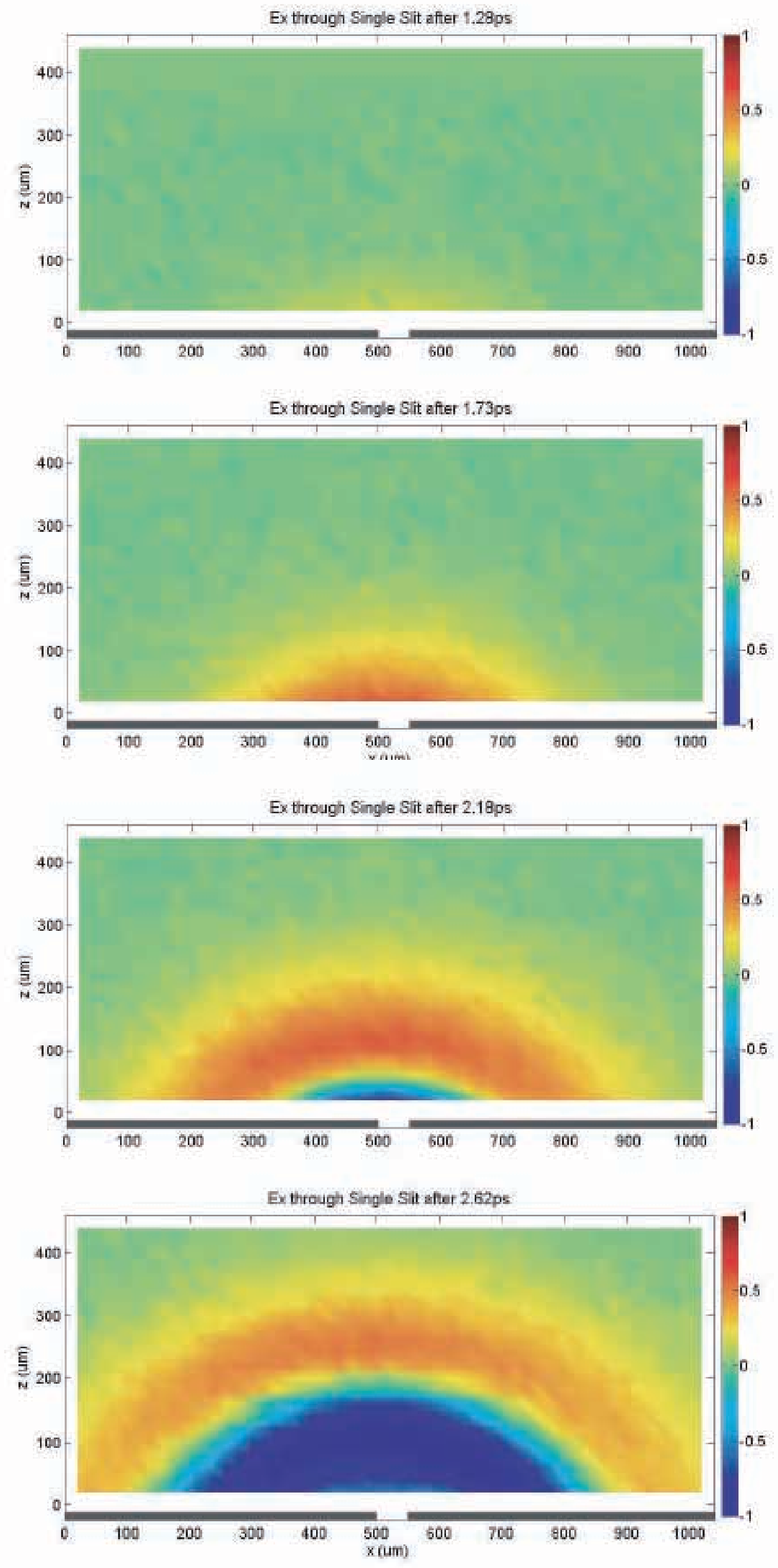}}}%
\subfigure[]{
\resizebox*{5cm}{!}{\includegraphics{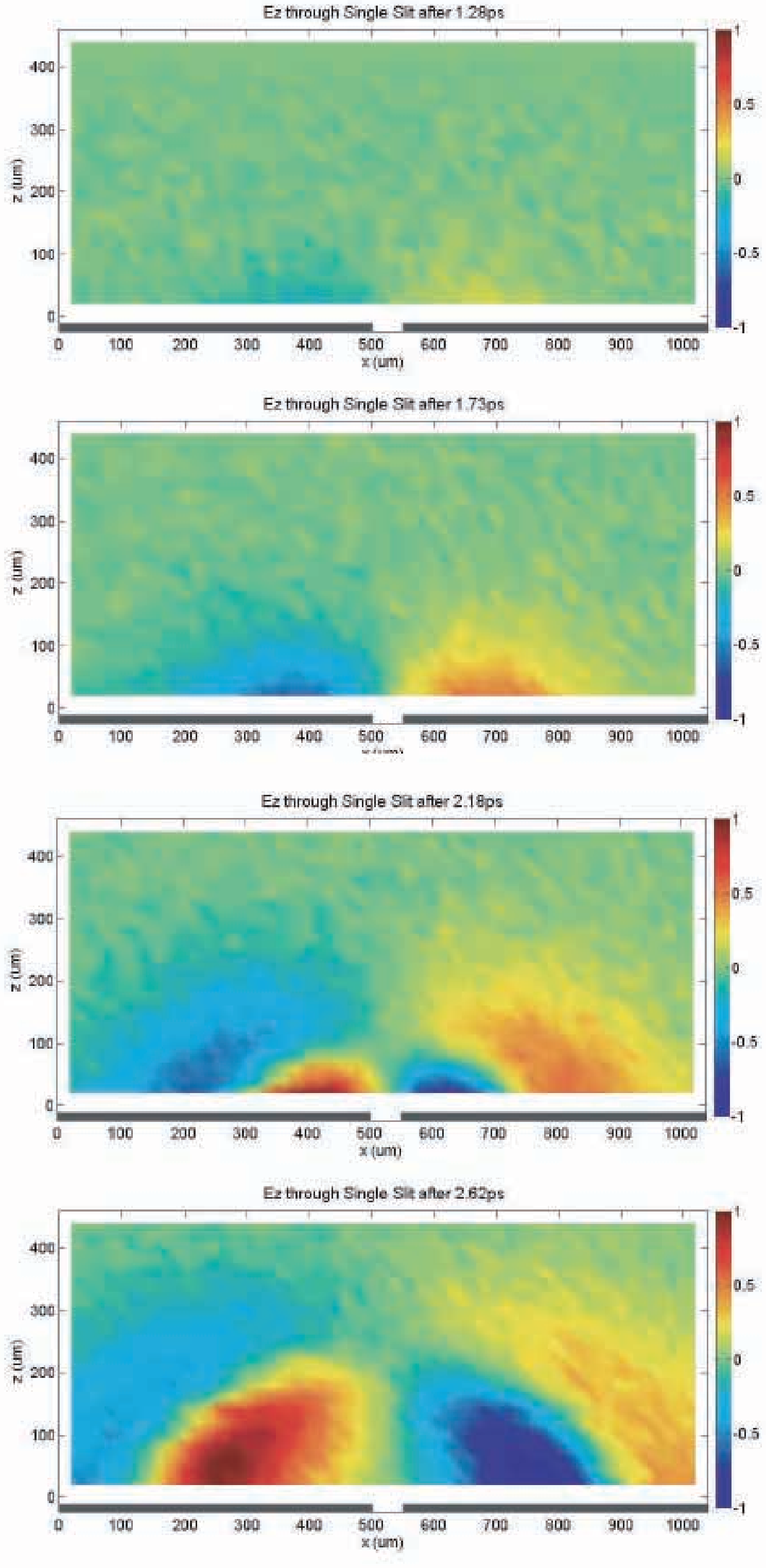}}}%
\caption{Electric field amplitude passing through a single slit:
(a) $E_x$ and (b) $E_z$ at $t=1.28 ps, 1.73 ps, 2.18 ps, 2.62 ps$.}%

\end{minipage}
\end{center}
\end{figure}

\begin{figure}
\begin{center}
\begin{minipage}{100mm}
\subfigure[]{
\resizebox*{5cm}{!}{\includegraphics{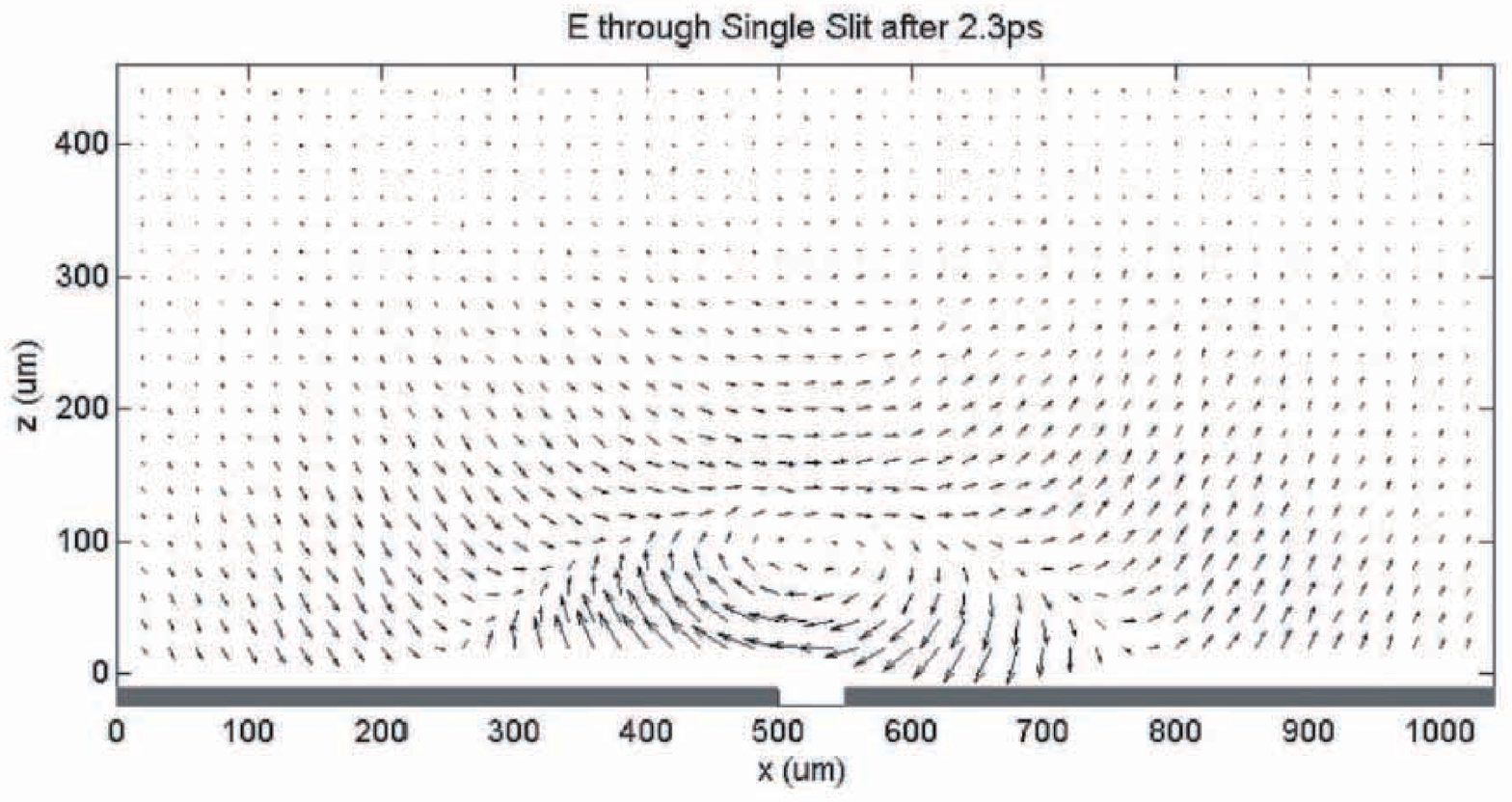}}}%
\caption{Electric field vector map. Arrows specify magnitude and direction of electric field vectors at
each positions.}%

\end{minipage}
\end{center}
\end{figure}

\subsection{Mapping electric and magnetic vecor fields using optical antennas}

Resonantly excited optical antennas generate strong local variations of
electric and magnetic vector fields at the nano scale. Most optical probes detect the intensity of light thus measuring
only the magnitude of vector fields. The direction of electric vector field was implicitly captured in the observation of single molecules
using NSOM where the orientation of each molecular dipole was determined \cite{Betzig2}. The longitudinal electric-field component of
Bessel beams was determined by using a polarization sensitive sheet of organic molecule as a probe \cite{Grosjean}.
Recently, we have shown that the explicit measurement of field vectors, both in magnitude and in direction, is possible \cite{Lee2}
when an optical antenna is properly combined with conventional probing techniques.
As illustrated in Fig. 8a, we attach a gold nanoparticle at the NSOM fibre tip and place it in the vicinity
of nanostructures. Then the local electric field drives the gold particle to act as an oscillating dipole
antenna. The radiated field intensity is measured after passing through a rotating polarizer. Since the dipole
radiation is critically dependent on the direction of induced dipole, which is in the same direction with
driving electric field, the rotating polarizer and subsequently the detection of varying intensity provides
information on the electric field vector. In this way, electric field vector maps have been demonstrated \cite{Lee2}
but without a full sense of the direction. Since the detection was made on the intensity of light passed through a
polarizer, it measures the direction modulo 180 degree.

In the terahertz frequency range, this difficulty can be
avoided through the terahertz time-domain spectroscopy, which is a spectroscopic technique
capable of measuring the phase and amplitude of the electric field in an ultra-wide THz
bandwidth \cite{Exter,Wu,Zhao}. A full vector field map has been obtained using the vectorial Fourier
transform two dimensional microscopy \cite{Seo} which utilizes the advantages of terahertz time domain spectroscopy
and the electro-optic sampling technique \cite{Gallot}.
In Fig. 9, we show the full electric field vector map of a THz pulse passing through a slit in a metal foil.
Time snapshots of the electric field vector components are given in the entire region of space behind the slits,
where red and blue colors represent positive and negative values of amplitude. This allows us to draw Faraday field lines
of electric field as given in Fig. 10.
Moreover, by taking the curl of measured electric field vector,  we obtain the time-dependent magnetic field
vector, and also the Poynting vector from the knowledge of electric and magnetic fields. In this way, a full vector field map
electromagnetic wave has been obtained with a resolution well below wavelength.

\section{Discussion}
Optical antenna is now becoming a major vehicle for nanophotonics research.
As a scaled down version of conventional radio and microwave antennas, optical antenna shares most of common
antenna characteristics. Optical analogues of monopole, dipole, directional and phased-array antennas have
been constructed and used for various nano-optics applications. We have seen that the resonance condition
of antenna changes significantly when we move toward the optical regime. The relatively low conductivity of
a metal in the optical domain invokes a collective motion of electrons, i.e. plasmons. Plasmonic resonance
is a new element in optical antenna systems. The old geometric resonance such as half wavelength
resonance is modified due to the excitation of plasmonic surface waves (surface plasmon polaritons)
with shortened wavelengths. The plasmonic resonance, arising in nanoparticles in the form of a dipole plasmon
resonance, proves that metal nanoparticle can be an efficient optical antenna.
Surface plasmon polaritons do not exist in the ideal case of a perfect conductor and become loosely bound
surfaces wave known as Sommerfeld-Zenneck waves\cite{Sommerfeld,Zenneck} for a good conductor such as
metal in the THz frequency domain \cite{Wang4,Jeon}. On the other hand, even for a perfect conductor, it was
shown that surface waves can exist if the surface is structured, e.g., with periodic grooves \cite{Goubau}.
Structures in a metal surface modifies the dispersion property of surface plasmon polaritons in a controlled
way \cite{Abajo2,Maier}. In antenna applications, grooves on the walls of a horn antenna were used to eliminate
the spurious diffractions at the edges of the aperture \cite{Kay}.
Structured surface can be applied to optical antennas as well \cite{Lezec}.
The combination of local plasmonic resonance and structural surface plasmon polaritons, which we may
simply term as plasmonics, provides ample possibilities for designing optical antennas.

Though we can learn a lot from the analogy between radio antennas and optical antennas, we can make deeper progress from
the difference. Since single light photon has much higher energy than a low frequency photon, through
field enhancement optical antennas can generate highly energetic local fields. This enables a new realm of
applications: active or nonlinear optical
processes such as optical antenna lasing and high harmonic generation, or communication with nanoscopic worlds such as
single molecules and nanobioimaging \cite{Garcia-Parajo}. In contrast to the wire type radio antennas, feeding currents
though transmission lines and controlling antenna by varying the feed gap are not easy in optical antennas.
This makes it difficult to have a circuit theory analysis of optical antenna. Nevertheless, efforts have been made to
develop a circuit theory for optical elements at nanoscales \cite{Engheta,EnghetaPRL2005,EnghetaPRL2008} and the
control of optical antennas have been investigated \cite{Merlein,Engheta2}.
Differences will increase as further progresses on optical antenna are being made.
Eventually optical antennas, as a beacon and a spear to
probe and control nanoworld, will develop into a major field of optics with exciting new discoveries waiting to be found.

\section*{Acknowledgement}
We thank D. S. Kim for invaluable discussions on the topic and J. H. Eberly for his encouragement.
This work is supported in part by KOSEF, KRF, MOCIE and the Seoul R $\&$ BD program


\begin{thebibliography}{10}
\markboth{Taylor \& Francis and I.T. Consultant}{Contemporary Physics}
\bibitem{Balanis} C. A. Balanis, {\em Antenna Theory: Analyses and Design}, John Wiley and Sons, Inc., Hoboken, New Jersey,
2005).

\bibitem{Frey1} H. G. Frey,  F. Keilmann, A. Kriele, R. Guckenberger, {\em Enhancing the resolution of scanning near-field optical microscopy
by a metal tip grown on an aperture probe}, Appl. Phys. Lett. 81, 5030 (2002).
\bibitem{Frey2} H. G. Frey, S. Witt, K. Felderer, and R. Guckenberger,{\em  High-resolution imaging of single fluorescent molecules
with the optical near-field of a metal tip}, Phys. Rev. Lett. 93, 200801 (2004).
\bibitem{Taminiau1} T. H. Taminiau, R. J. Moerland, F. B. Segerink, L. Kuipers, and N. F. van Hulst,
{\em $\lambda /4$ resonance of an optical monopole antenna probed by single molecule fluorescence}, Nano Lett. 7, 28 (2007).
\bibitem{Taminiau2} T. H. Taminiau, F. B. Segerink, and N. F. van Hulst, "A Monopole Antenna at Optical Frequencies: Single-
Molecule Near-Field Measurements," IEEE Trans. Antennas Propag. 55, 3010-3017 (2007).
\bibitem{Taminiau3} T. H. Taminiau, F. B. Segerink, R. J. Moerland, L. Kuipers, and N. F. van Hulst, "Near-field driving of a
optical monopole antenna," J. Opt. A: Pure Appl. Opt. 9, S315-S321 (2007).


\bibitem{Grober} R. D. Grober, R. J. Schoelkopf, and D. E. Prober, {\em Optical antenna: Towards a unity efficiency
near-field optical probe}, Appl. Phys. Lett. 70, 1354-1356, (1997).

\bibitem{Lewis} A. Lewis, M. Isaacson, A. Harootunian, and A. Murray, {\em Development of a 500 A spatial resolution light microscope.
I. Light is efficiently transmitted through $\lambda /16$ diameter apertures}, Ultramicroscopy, 13, 227 (1984)
\bibitem{Pohl} D.W. Pohl, W. Denk, and M. Lanz, {\em Optical stethoscopy: Image recording with resolution $\lambda /20$},
Appl. Phys. Lett., 44, 651 (1984).
\bibitem{Betzig} E. Betzig, J. K. Trautman, T. D. Harris, J. S. Weiner, and R. L. Kostelak,
{\em Breaking the Diffraction Barrier: Optical Microscopy on a Nanometric Scale}, Science 251, 1468 (1991).

\bibitem{Muhlschlegel} P. M\"{u}hlschlegel, H. J. Eisler, O. J. F. Martin, B. Hecht, and D. W. Pohl, {\em Resonant Optical Antennas},
Science 308, 1607 (2005).
\bibitem{Ghenuche} P. Ghenuche, S. Cherukulappurath, T. H. Taminiau, N. F. van Hulst, and R. Quidant,
{\em Spectroscopic Mode Mapping of Resonant Plasmon Nanoantennas}, Phys. Rev. Lett. 101, 116805 (2008)

\bibitem{Schuck} P. J. Schuck, D. P. Fromm, A. Sundaramurthy, G. S. Kino, and W. E. Moerner,
{\em Improving the Mismatch between Light and Nanoscale Objects with Gold Bowtie Nanoantennas}, Phys. Rev. Lett. 94, 017402 (2005).

\bibitem{Farahani1} J. N. Farahani, D.W. Pohl, H.-J. Eisler, and B. Hecht
{\em Single quantum dot coupled to a scanning optical antenna: a tunable superemitter}, Phys. Rev. Lett. 95, 017402 (2005).
\bibitem{Farahani2} J. N. Farahani, H.-J. Eisler, D.W. Pohl, M. Pavius, P. Fl\"{u}ckiger, P. Gasser, and B. Hecht,
{\em Bow-tie optical antenna probes for single-emitter scanning near-field optical microscopy}, Nanotechnology 18, 125506 (2007).
\bibitem{Guo} H. Guo, T. P. Meyrath, T. Zentgraf, N. Liu, L. Fu, H. Schweizer, and H. Giessen,
{\em Optical resonances of bowtie slot antennas and their geometry and material dependence}, Opt. Express 16, 7756-7766 (2008).
\bibitem{Fischer} H. Fischer and O. J. F. Martin, {\em Engineering the optical response of plasmonic nanoantennas},
Opt. Express 16, 9144-9154 (2008).
\bibitem{Huang} C. Huang, A. Bouhelier, G. Colas des Francs, A. Bruyant, A. Guenot, E. Finot, J.-C. Weeber, and A. Dereux
{\em Gain, detuning, and radiation patterns of nanoparticle optical antennas}, Phys. Rev. B 78, 155407 (2008).
\bibitem{Fromm} D. P. Fromm, A. Sundaramurthy, P. J. Schuck, G. Kino, and W. E. Moerner, {\em Gap-Dependent Optical Coupling
of Single ¡±Bowtie¡± Nanoantennas Resonant in the Visible}, Nano Lett. 4, 957-961 (2004).
\bibitem{Wang} L. Wang, S. M. Uppuluri, E. X. Jin, and X. Xu, {\em Nanolithography Using High Transmission Nanoscale Bowtie
Apertures}, Nano Lett. 6, 361-364 (2006).
\bibitem{Sundaramurthy} A. Sundaramurthy, P. J. Schuck, N. R. Conley, D. P. Fromm, G. S. Kino, andW. E.Moerner, {\em Toward Nanometer-
Scale Optical Photolithography: Utilizing the Near-Field of Bowtie Optical Nanoantennas}, Nano Lett. 6, 355-360 (2006).


\bibitem{Kuhn} S. K\"{u}hn, U. Hakanson, L. Rogobete, and V. Sandoghdar, "Enhancement of Single-Molecule Fluorescence
Using a Gold Nanoparticle as an Optical Nanoantenna," Phys. Rev. Lett. 97, 017402 (2006).
\bibitem{Bharadwaj1} P. Bharadwaj, P. Anger and L. Novotny, {\em Nanoplasmonic enhancement of single-molecule fluorescence},
Nanotechnology 18, 044017 (2007).
\bibitem{Bharadwaj2} P. Bharadwaj and L. Novotny, {\em Spectral dependence of single molecule fluorescence enhancement}, Optics Express
15, 14266-14274 (2007).
\bibitem{Tam} F. Tam, G. P. Goodrich, B. R. Johnson, and N. J. Halas, {\em Plasmonic Enhancement of Molecular Fluorescence},
Nano Lett. 7, 496-501 (2007).
\bibitem{Bakker} R. M. Bakker, H.-K. Yuan, Z. Liu, V. P. Drachev, A. V. Kildishev, V. M. Shalaev, R. H. Pedersen, S. Gresillon,
A. Boltasseva, {\em Enhanced localized fluorescence in plasmonic nanoantennae}, Appl. Phys. Lett. 92, 043101 (2008).

\bibitem{Gersten} J. Gersten and A. Nitzan, {\em Electromagnetic theory of enhanced Raman scattering by molecules adsorbed on rough surfaces}, J. Chem. Phys. 73, 3023-3037 (1980).
\bibitem{Nie} S. Nie, and S. R. Emory, ¡±Probing Single Molecules and Single Nanoparticles by Surface-Enhanced Raman
Scattering,¡± Science 275, 1102-1106 (1997).
\bibitem{Kneipp} K. Kneipp, Y. Wang, H. Kneipp, L. T. Perelman, I. Itzkan, R. R. Dasari, and M. S. Feld, ¡±Single Molecule
Detection Using Surface-Enhanced Raman Scattering (SERS),¡± Phys. Rev. Lett. 78, 1667 (1997).
\bibitem{Xu} H. Xu, E. J. Bjerneld, M. K¡§all, and L. Brjesson, ¡±Spectroscopy of Single Hemoglobin Molecules by Surface
Enhanced Raman Scattering,¡± Phys. Rev. Lett. 83, 4357 (1999).
\bibitem{Felidj} N. Felidj, J. Aubard, G. Levi, J. R. Krenn, A. Hohenau, G. Schider, A. Leitner, and F. R. Aussenegg, ¡±Optimized
surface-enhanced Raman scattering on gold nanoparticle arrays,¡± Appl. Phys. Lett. 82, 3095-3097 (2003).
\bibitem{Rogobete} L. Rogobete, F. Kaminski, M. Agio, and V. Sandoghdar, ¡±Design of plasmonic nanoantennae for enhancing
spontaneous emission,¡± Opt. Lett. 32, 1623-1625 (2007).
\bibitem{Taminiau} T. H. Taminau, F. D. Stefani, F. B. Segerink and N. F. Van Hulst, ¡±Optical antennas direct single-molecule
emission,¡± Nature Photonics 2, 234-237 (2008).

\bibitem{Garcia-Parajo} M. F. Garcia-Parajo, {\em  Optical antennas focus in on biology}, Nature Photon. 2, 201-203 (2008).

\bibitem{Imura} K. Imura, T. Nagahara, and H. Okamoto, {\em Near-field optical imaging of plasmon modes in gold nanorods}, J. Chem. Phys. 122,
154701 (2005).
\bibitem{Novotny} L. Novotny, {\em Effective wavelength scaling for optical antennas}, Phys. Rev. Lett. 98, 266802 (2007).
\bibitem{Hanson} G. W. Hanson, {\em On the Applicability of the Surface Impedance Integral Equation for Optical and Near Infrared
Copper Dipole Antennas}, IEEE Trans. Ant. and Prop., vol 54, 3677-3685 (2006).
\bibitem{Ditlbacher} H. Ditlbacher, A. Hohenau, D. Wagner, U. Kreibig, M. Rogers, F. Hofer, F. R. Aussenegg, and J. R. Krenn,
{\em Silver Nanowires as Surface Plasmon Resonators}, Phys. Rev. Lett. 95, 257403 (2005).
\bibitem{Neubrech} F. Neubrech {\it et al.}, {\em Resonances of individual metal nanowires in the infrared}, Appl. Phys. Lett. 89, 253104 (2006).
\bibitem{Laroche} T. Laroche and C. Girard, {\em Near-field optical properties of single plasmonic nanowires}, Appl. Phys. Lett. 89, 233119 (2006).
\bibitem{Payne} E. K. Payne, K. L. Shuford, S. Park, G. C. Schatz, and C. A. Mirkin, {\em Multipole Plasmon Resonances in Gold Nanorods},
J. Phys. Chem. B 110, 2150-2154, (2006).
\bibitem{Crozier} K. B. Crozier, A. Sundaramurthy, G. S. Kino, and C. F. Quate, {\em Optical antennas: Resonators for local field enhancement}, J. Appl. Phys. 94, 4632-4642, (2003).
\bibitem{Schider} G. Schider, J. R. Krenn, A. Hohenau, H. Ditlbacher, A. Leitner, F. R. Aussenegg, W. L. Schaich, I. Puscasu, B.
Monacelli, and G. Boreman, {\em Plasmon dispersion relation of Au and Ag nanowires}, Phys. Rev. B 68, 155427
(2003).

\bibitem{Ritchie} R. H. Ritchie,  {\em  Plasma losses by fast electrons in thin films}, Phys. Rev. 106, 874.881
(1957).
\bibitem{Raether} H. Raether, {\em Surface Plasmons on Smooth and Rough Surfaces and on Gratings},
Springer, Berlin, (1988).
\bibitem{Kelly} K. L. Kelly, E. Coronado, L. L. Zhao, and G. C. Schatz,
{\em The Optical Properties of Metal Nanoparticles: The Influence of Size, Shape, and Dielectric Environment},
J. Phys. Chem. B 107, 668-677 (2003).

\bibitem{Barnes} W. L. Barnes, , A. Dereux, and T. W. Ebbesen,  {\em  Surface plasmon subwavelength optics}, Nature 424, 824.830 (2003).
\bibitem{SAMaier} S. A. Maier and H. A. Atwater, {\em Plasmonics: Localization and guiding of electromagnetic
energy in metal/dielectric structures}, J. Appl. Phys. 98, 011101 (2005).
\bibitem{Lal} S. Lal, S. Link, and N. J. Halas, {\em Nano-optics from sensing to waveguiding}, Nature Photonics 1, 641-648 (2007).



\bibitem{Genet} C. Genet and T. W. Ebbesen, {\em Light in tiny holes}, Nature, 445, 39-46 (2007).
\bibitem{Bethe} H. A. Bethe, {\em Theory of diffraction by small holes}, Phys. Rev. 66, 163-182 (1944).
\bibitem{Bouwkamp} C. J. Bouwkamp, {\em Diffraction theory}, Reports on Progress in Physics XVIII, 35 (1954).
\bibitem{Ebbesen} T. W. Ebbesen, H. J. Lezec, H. F. Ghaemi,  T. Thio, and P. A. Wolff, {\em Extraordinary
optical transmission through sub-wavelength hole arrays}, Nature 391, 667-669 (1998).

\bibitem{Degiron1} A. Degiron, H. J. Lezec, N. Yamamoto, and T. W. Ebbesen, {\em Optical transmission properties of a single
subwavelength aperture in a real metal}, Optics Commun. 239, 61-66 (2004).
\bibitem{Degiron2} A. Degiron and T. W. Ebbesen, {\em Analysis of the transmission process through single apertures surrounded
by periodic corrugations}, Opt. Express 12, 3694-3700 (2004).
\bibitem{Zakharian} A. R. Zakharian, M. Mansuripur, and J. V. Moloney, {\em Transmission of light through small elliptical
apertures}, Opt. Express 12, 2631-2648 (2004).
\bibitem{Alaverdyan} Y. Alaverdyan, B. Sepulveda, L. Eurenius, E. Olsson, and M. Kall, {\em Optical antennas based on coupled
nanoholes in thin metal films}, Nature Phys. 3, 884-889 (2007).
\bibitem{Rindzevicius} T. Rindzevicius, Y. Alaverdyan, B. Sepulveda, T. Pakizeh, M. Kall, R. Hillenbrand, J. Aizpurua, and F. J.
Garcia de Abajo, {\em Nanohole plasmons in optically thin gold films}, J. Phys. Chem. C 111, 1207 (2007).
\bibitem{Prikulis} J. Prikulis, P. Hanarp, L. Olofsson, D. Sutherland, and M. Kall, {\em Optical spectroscopy of nanometric holes in
thin gold films}, Nano Lett. 4, 1003-1007 (2004).
\bibitem{Sepulveda} B. Sepulveda, Y. Alaverdyan, J. Alegret, M. Kall and P. Johansson, {\em Shape effects in the localized surface plasmon
resonance of single nanoholes in thin metal films}, Opt. Express 16, 5609-5616 (2008).
\bibitem{Popov} E. Popov, M. Neviere, A. Sentenac, N. Bonod, A.-L. Fehrembach, J. Wenger, P.-F. Lenne, and H. Rigneault,
{\em Single-scattering theory of light diffraction by a circular subwavelength aperture in a finitely conducting screen},
J. Opt. Soc. Am. A 24, 339-358, (2007).
\bibitem{Abajo} F. J. Garcia de Abajo, {\em Light transmission through a single cylindrical hole in a metallic film},
Opt. Express 10, 1475-1484 (2002).


\bibitem{Shi} X. Shi, L. Hesselink, and R. L. Thornton, {\em Ultrahigh light transmission through a C-shaped nanoaperture}, Opt.
Lett. 28, 1320-1322 (2003).
\bibitem{Matteo} J. A. Matteo, D. P. Fromm, Y. Yuen, P. J. Schuck, W. E. Moerner, and L. Hesselink, {\em Spectral analysis of strongly
enhanced visible light transmission through single C-shaped nanoapertures}, Appl. Phys. Lett. 85, 648 (2004).
\bibitem{Sun} L. Sun and L. Hesselink, {\em Low-loss subwavelength metal C-aperture waveguide}, Opt. Lett. 31, 3606 (2006).
\bibitem{Lee} J. W. Lee, M. A. Seo, D. J. Park and D. S. Kim, S. C. Jeoung, C. Lienau, Q. H. Park, P. C. M. Planken, {\em Shape resonance omni-directional terahertz
filters with near-unity transmittance}, Opt. Express 14, 1253-1259 (2006).


\bibitem{Jin} E. X. Jin and X. Xu, {\em Plasmonic effects in near-filed optical transmission enhancement through a single bowtieshaped
aperture}, Appl. Phys. B. 84, 3 (2006).
\bibitem{Ishihara} K. Ishihara, K. Ohashi, T. Ikari, H. Minamide, H. Yokoyama, J. Shikata, and H. Ito, {\em Terahertz-wave near-field
imaging with subwavelength resolution using surface-wave-assistanted bow-tie aperture}, Appl. Phys. Lett. 89, 201120 (2006).
\bibitem{Wang1} L. Wang, S. M. Uppuluri, E. X. Jin, and X. Xu, {\em Nanolithography using high transmission nanoscale bowtie
apertures}, Nano Lett. 6, 361 (2006).
\bibitem{Wang2} L.Wang and X. Xu, {\em High transmission nanoscale bowtie-shaped aperture probe for near-field optical imaging},
Appl. Phys. Lett. 90, 261105 (2007).

\bibitem{Garcia-Vidal} F. J. Garcia-Vidal, E. Moreno, J. A. Porto, and L. Martin-Moreno, {\em Transmission
of Light through a Single Rectangular Hole}, Phys. Rev. Lett. 95, 193901 (2005).
\bibitem{DSKim} M. A. Seo, A. J. L. Adam, J. H. Kang, J. W. Lee,  K. J. Ahn,  Q. H. Park,
P. C. M. Planken, and D. S. Kim, {\em  Near field imaging of terahertz focusing onto rectangular apertures},
Opt. Express 16, 20484-20489 (2008).

\bibitem{Zhang} Z. J. Zhang, R. W. Peng, Z. Wang, F. Gao, X. R. Huang, W. H. Sun, Q. J. Wang,
and Mu Wang, {\em Plasmonic antenna array at optical frequency made by nanoapertures}, Appl. Phys. Lett. 93, 171110 (2008).

\bibitem{Lezec} H. J. Lezec, A. Degiron, E. Devaux, R. A. Linke, L. Martin-Moreno, F. J. Garcia-Vidal, and T. W. Ebbesen,
{\em  Beaming light from a subwavelength aperture}, Science 297, 820-822 (2002).
\bibitem{Degiron}  A. Degiron, and T. W. Ebbesen, {\em Analysis of the transmission process through single
apertures surrounded by periodic corrugations}, Opt. Express 12, 3694-3700 (2004).
\bibitem{Martin-Moreno} L. Martin-Moreno, F. J. Garcia-Vidal, H. J. Lezec, A. Degiron, and T. W. Ebbesen,
{\em Theory of highly directional emission from a single subwavelength aperture surrounded by surface corrugations},
Phys. Rev. Lett. 90, 167401 (2003).

\bibitem{Yagi} H. Yagi, {\em Beam transmission of ultra short waves}, Proc. IRE 16, 715-741 (1928).
\bibitem{Taminiau4} T. H. Taminiau, F. D. Stefani, and N. F. van Hulst, {\em Enhanced directional excitation and emission
of single emitters by a nano-optical Yagi-Uda antenna}, Opt. Express 16, 10858-10866 (2008).
\bibitem{Li} J. Li, A. Salandrino, and N. Engheta, {\em Shaping light beams in the nanometer scale: A Yagi-Uda nanoantenna
in the optical domain}, Phys. Rev. B 76, 245403-245407 (2007).
\bibitem{Hofmann} H. F. Hofmann, T. Kosako, and Y. Kadoya,{\em Design parameters for a nano-optical Yagi-Uda antenna}, New
J. Phys. 9, 217-217 (2007).


\bibitem{Bozhevolnyi1} T. S$\o$ndergaard and S. I. Bozhevolnyi, {\em  Slow-plasmon resonant nanostructures: scattering and field enhancements},
Phys. Rev. B 75, 073402 (2007).
\bibitem{Bozhevolnyi2} T. S$\o$ndergaard, J. Beermann, A. Boltasseva, and S. I. Bozhevolnyi, {\em Slow-plasmon resonant-nanostrip
antennas: analysis and demonstration}, Phys. Rev. B 77, 115420 (2008)



\bibitem{Cubukcu} E. Cubukcu, E. A. Kort, K. B. Crozier, and F. Capasso, {\em Plasmonic laser antenna}, Appl. Phys. Lett. 89, 093120 (2006).
\bibitem{Yu} N. Yu, E. Cubukcu, L. Diehl, M. A. Belkin, K. B. Crozier, and F. Capasso, D. Bour, S. Corzine, and G. Hoflerer,
{\em  Plasmonic quantum cascade laser antenna}, Appl. Phys. Lett. 91, 173113 (2007).
\bibitem{Kim} S. Kim, J. Jin, Y.-J. Kim, I.-Y. Park, Y. Kim, and S.-W. Kim, {\em High-harmonic generation by resonant plasmon
field enhancement}, Nature 453, 757-760, (2008).


\bibitem{CRC} M. J. Weber, {\em Handbook of Optical Materials}, CRC Press, London (2003).


\bibitem{Kreibig} U. Kreibig and M. Vollmer, {\em Optical Properties of Metal Clusters}, Springer, Berlin, (1995).
\bibitem{Bohren} C. F. Bohren and D. R. Huffman, {\em Absorption and Scattering of Light by Small Particles}, Wiley, New York, (1983).



\bibitem{MMeier} M. Meier and A. Wokaun, {\em Enhanced fields on large metal particles: dynamic depolarization}, Opt. Lett. 8, 581 (1983).
\bibitem{Wokaun} A. Wokaun, J. P. Gordon, and P. F. Liao, {\em  Radiation Damping in Surface- Enhanced Raman Scattering},
Phys. Rev. Lett. 48, 957 (1982).

\bibitem{Mie} G. Mie, Ann. Phys. 25, 377 (1908).

\bibitem{Porto} J. A. Porto, F. J. Garcia-Vidal, and J. B. Pendry, {\em Transmission resonances on metallic gratings with very narrow
slits}, Phys. Rev. Lett. 83, 2845 (1999).
\bibitem{Treacy} M. M. J. Treacy, {\em Dynamical diffraction explanation of the anomalous transmission of light through metallic
gratings}, Phys. Rev. B 66, 195105 (2002).
\bibitem{Cao} Q. Cao and P. Lalanne, {\em Negative role of surface plasmons in the transmission of metallic gratings with very narrow slits}, Phys. Rev. Lett. 88, 057403 (2002).
\bibitem{Lee1} K. G. Lee, and Q-Han Park, {\em Coupling of surface plasmon polaritions and light in metallic nanoslits}, Phys.
Rev. Lett. 95, 103902 (2005).


\bibitem{Seo2} M. A. Seo {\it et al.} {\em Terahertz nanogap device for field enhancement}, submitted.
\bibitem{QPark} J. H. Kang, D. S. Kim, and Q. H. Park, {\em Local capacitor model for plasmonic electric field enhancement}, submitted.

\bibitem{Betzig2} E. Betzig and R. J. Chichester, {\em Single molecules observed by near-field scanning optical microscopy},
Science 262, 1422-1425 (1993).
\bibitem{Grosjean} T. Grosjean and D. Courjon, {\em Photopolymers as vectorial sensors of the electric field}, Opt. Express
14, 2203-2210 (2006).
\bibitem{Lee2} K. G. Lee et {\it al.}, {\em Vector field microscopic imaging of light}, Nature Photon. 1, 53.56 (2007).
\bibitem{Zenhausern}  F. Zenhausern,, M. P. O'Boyle, and H. K. Wickramasinghe, {\em Apertureless near-field optical
microscope}, Appl. Phys. Lett. 65, 1623.1625 (1994).
\bibitem{Kawata} S. Kawata and Y. Inouye, {\em  Scanning probe optical microscopy using a metallic probe tip},
Ultramicroscopy 57, 313.317 (1995).

\bibitem{Seo} M. A. Seo, A. J. L. Adam, J. H. Kang, J. W. Lee, S. C. Jeoung, Q. H. Park,
P. C. M. Planken, and D. S. Kim, {\em Fourier-transform terahertz near-field imaging of one-dimensional slit arrays: mapping of
electric-field-, magnetic-field-, and Poynting vectors}, Opt. Express 15, 11781-11789 (2007).


\bibitem{Exter} M. van Exter and D. R. Grischkowsky, ¡°Characterization of an optoelectronic terahertz beam system,¡± IRE
Trans. Microwave Theory and Tech. 38, 1684-1691 (1990).
\bibitem{Wu} Q. Wu and X. C. Zhang, ¡°Free-space electro-optic sampling of terahertz beams,¡± Appl. Phys. Lett. 67,
3523-3525 (1995).
\bibitem{Zhao} G. Zhao, R. N. Schouten, N. van der Valk, W. T. Wenckebach, and P. C. M. Planken, ¡°Design and
performance of a THz emission and detection setup based on a semi-insulating GaAs emitter,¡± Rev. Sci.
Instrum. 73, 1715-1719 (2002).



\bibitem{Gallot} G. Gallot and D. R. Grischkowsky, ¡°Electro-optic detection of terahertz radiation,¡± J. Opt. Soc. Am. B 16,
1204-1212 (1999).

\bibitem{Sommerfeld} A. Sommerfeld, Ann. Phys. U. Chemie 67-1,233, (1899).
 \bibitem{Zenneck} J. Zenneck, Ann. d. Phys. 23, 846, (1907).
 \bibitem{Wang4} K. Wang and D.M.Mittleman, {\em Metal wires for THz wave guiding},  Nature 432, 376 (2004).
\bibitem{Jeon} T.I. Jeon, J. Zhang, D. Grischkowsky, {\em THz Sommerfeld wave propagation on a single metal wire},
 Appl. Phys. Lett. 86, 161904 (2005),
 \bibitem{Goubau} G. Goubau, {\em Surface waves and their application to transmission lines}, J. Appl. Phys. 21, 1119 (1950).
\bibitem{Abajo2} F. J. Garcia de Abajo, and J. J. Sa\'{e}nz,  {\em  Electromagnetic Surface Modes in Structured Perfect-Conductor Surfaces},
Phys. Rev. Lett. 95, 233901 (2005).
\bibitem{Maier} S. A. Maier, S. R. Andrews, L. Martin-Moreno,and F. J. Garcia-Vidal,
{\em Terahertz Surface Plasmon-Polariton Propagation and Focusing on Periodically Corrugated Metal Wires}, Phys. Rev. Lett. 97, 176805 (2006).

\bibitem{Kay} A. F. Kay, {\em The scalar feed}, AFCRL Rep. 64-347, AD601609, March 1964.

\bibitem{Engheta} N. Engheta, {\em Circuits with light at nanoscales: optical nanocircuits inspired by metamaterials} Science 317, 1698 (2007);
\bibitem{EnghetaPRL2005} N. Engheta, A. Salandrino, and A. Al\`{u},
{\em Circuit elements at optical frequencies: nanoinductors, nanocapacitors, and nanoresistors}, Phys. Rev. Lett. 95, 095504 (2005)
\bibitem{EnghetaPRL2008} A. Al\`{u} and N. Engheta, {\em Input impedance, nanocircuit loading, and radiation tuning of optical
nanoantennas}, Phys. Rev. Lett. 101, 043901 (2008)
\bibitem{Burke} P. J. Burke, S. Li, and Z. Yu, {\em Quantitative Theory of Nanowire and Nanotube Antenna Performance},  IEEE Trans. Nanotech. 5, 314-334, (2006)


\bibitem{Merlein} J. Merlein et al. {\em Nanomechanical control of an optical antenna}, Nature Photon. 2, 230-233 (2008).
\bibitem{Engheta2} A. Al\`{u} and N. Engheta, {\em Tuning the scattering response of optical
nanoantennas with nanocircuit loads}, Nature Photon. 2, 307-310 (2008).



\end{thebibliography}
\end{document}